\documentclass[11pt,a4paper]{article}
\pdfoutput=1
\usepackage{jheppub}
\usepackage[utf8]{inputenc}
\usepackage{epstopdf}
\usepackage{tabularx}
\usepackage{amsmath,amssymb, amsfonts,amsbsy,mathrsfs,graphicx, booktabs, slashed, xcolor, adjustbox}
\usepackage{latexsym}
\usepackage{hyperref}
\usepackage{bm}
\usepackage{comment}
\usepackage{bbm}
\usepackage{scalerel}

\newcommand{\F}{\mathsf{F}}
\newcommand{\trF}[1]{\F[#1]}
\newcommand{\trFt}[1]{\F[#1]}
\newcommand{\G}{\mathsf{G}}

\def\equationautorefname~#1\null{Eq.\,(#1)\null}
\def\sectionautorefname~#1\null{Sec.\,#1\null}
\def\subsectionautorefname~#1\null{Sec.\,#1\null}
\def\figureautorefname~#1\null{Fig.\,#1\null}
\def\appendixautorefname~#1\null{App.\,#1\null}

\graphicspath{{figs/}}

\usepackage{tikz}
\usepgflibrary{shapes.misc,shapes.geometric,arrows.meta}
\usetikzlibrary{positioning, decorations.markings, calc}
\newcommand{\minitab}[2][c]{\begin{tabular}{@{}#1@{}}#2\end{tabular}}

\usepackage{ifpdf,feynmp}
\ifpdf\fi
\AtBeginDocument{\begin{fmffile}{figs/fgraph}
\fmfcmd{%
thin := .5mm;
thick := .7mm;
arrow_ang := 11;
arrow_len := 3.5thick;
curly_len := 2.5thick;
color myg,myp,myr, Dsix, Deight;
Dsix := .9blue;
Deight := .5green;
myg := .7green;
myp := .7blue;
myr := .7red;}
}
\newcommand{\Dsix}[1]{{\color{blue!90!black}#1}}
\newcommand{\Deight}[1]{{\color{green!50!black}#1}}

\makeatletter\g@addto@macro\bfseries{\boldmath}\makeatother

\newcommand{\Tr}{\mathrm{Tr}}

\newcommand{\cref}[1]{Chapter~\ref{ch:#1}}

\newcommand{\nn}{\nonumber }

\newcommand{\beq}{\begin{equation}} 
\newcommand{\eeq}{\end{equation}} 
\newcommand{\ba}{\begin{array}}  
\newcommand{\ea}{\end{array}} 
\newcommand{\bea}{\begin{eqnarray}}  
\newcommand{\eea}{\end{eqnarray} }  
\newcommand{\be}{\begin{eqnarray}}  
\newcommand{\ee}{\end{eqnarray} }  
\newcommand{\bal}{\begin{align}}
\newcommand{\eal}{\end{align}}   
\newcommand{\ben}{\begin{enumerate}}  
\newcommand{\een}{\end{enumerate}}  
\newcommand{\bc}{\begin{center}}
\newcommand{\ec}{\end{center}} 
\newcommand{\bt}{\begin{table}}
\newcommand{\et}{\end{table}}  
\newcommand{\btb}{\begin{tabular}}
\newcommand{\etb}{\end{tabular}}

\renewcommand{\[}{\left[}
\renewcommand{\]}{\right]}
\renewcommand{\(}{\left(}
\renewcommand{\)}{\right)}
\def\bes{\begin{equation*}}
\def\ees{\end{equation*}}
\def\bead{\begin{aligned}}
\def\eead{\end{aligned}}
\def\bmat{\left(\begin{matrix}}
\def\emat{\end{matrix}\right)}

\newcommand{\cO}{{\mathcal O}} 
\newcommand{\co}{{\mathcal O}} 
\newcommand{\cL}{{\mathcal L}}

\newcommand{\cA}{{\mathcal A}}   

\newcommand\ptwiddle[1]{\mathord{\mathop{#1}\limits^{\scriptscriptstyle(\sim)}}}

\author[a,b]{Quentin Bonnefoy,}
\author[c,d]{Gauthier Durieux,}
\author[e,f]{Jasper Roosmale Nepveu}

\affiliation[a]{Berkeley Center for Theoretical Physics, Department of Physics, University of California,\\ Berkeley, CA 94720, USA}
\affiliation[b]{Theoretical Physics Group, Lawrence Berkeley National Laboratory, Berkeley, CA 94720, USA}
\affiliation[c]{CERN, Theoretical Physics Department, Geneva 23 CH-1211, Switzerland}
\affiliation[d]{Centre for Cosmology, Particle Physics and Phenomenology, Universit\'e catholique de Louvain, 1348 Louvain-la-Neuve, Belgium}
\affiliation[e]{Humboldt-Universit\"at zu Berlin, Institut f\"ur Physik, Newtonstr.\ 15, 12489 Berlin, Germany}
\affiliation[f]{Deutsches Elektronen-Synchrotron DESY, Notkestr.\ 85, 22607 Hamburg, Germany}
\emailAdd{q.bonnefoy@berkeley.edu}
\emailAdd{gauthier.durieux@uclouvain.be}
\emailAdd{jasper.roosmalenepveu@physik.hu-berlin.de}

\title{%
Higher-derivative relations between scalars and gluons
}
\abstract{%
We extend the covariant color-kinematics duality introduced by Cheung and Mangan to effective field theories.
We focus in particular on relations between the effective field theories of gluons only and of gluons coupled to bi-adjoint scalars.
Maps are established between their respective equations of motion and between their tree-level scattering amplitudes.
An additional rule for the replacement of flavor structures by kinematic factors realizes the map between higher-derivative amplitudes.
As an example of new relations, the pure-gluon amplitudes of mass dimension up to eight, featuring insertions of the $F^3$ and $F^4$ operators which satisfy the traditional color-kinematics duality, can be generated at all multiplicities from just renormalizable amplitudes of gluons and bi-adjoint scalars.
We also obtain closed-form expressions for the kinematic numerators of the dimension-six gluon effective field theory, which are valid in $D$ space-time dimensions.
Finally, we find strong evidence that this extended covariant color-kinematics duality relates the $(DF)^2+$YM$(+\phi^3)$ theories which, at low energies, generate infinite towers of operators satisfying the traditional color-kinematics duality, beyond aforementioned $F^3$ and $F^4$ ones.
}

\begin{document}
\begin{flushright}
CERN-TH-2023-174
\\
IRMP-CP3-23-58
\\
DESY-23-145
\\
HU-EP-23/53-RTG
\end{flushright}
\flushbottom
%\sloppy %https://tex.stackexchange.com/questions/9107/how-can-i-make-my-text-never-go-over-the-right-margin-by-always-hyphenating-or-b

\makeatletter\renewcommand{\@fpheader}{\ }\makeatother
\maketitle

%%%%%%%%%%%%%%%%%%%%%%%%%%%%%%%%%%%%%%%%%%%%%%%%%%%%%%%%%%%%%%%%%%%%%%%%%%%%%%%%%%%%%%%
\section{Introduction}

Scattering amplitudes are of wide interest in high-energy physics for their close connection to observables and for their remarkable mathematical properties.
Their study has led to new fascinating discoveries, some of which are not at all obvious from the perspective of Lagrangians or Feynman rules.
One prominent example of such unexpected structure is the relation between Yang-Mills (YM) and gravity theories, originally identified by Kawai, Lewellen and Tye (KLT) as a mapping from open to closed string amplitudes~\cite{Kawai:1985xq}.
In its low-energy limit, this relation enables the calculation of tree-level graviton amplitudes from the product of two, arguably simpler, gluon amplitudes convoluted with a matrix of kinematic functions dubbed the KLT kernel.
The basis independence of these double-copy relations relies on the low-energy limit of string monodromy relations.

It was later found by Bern, Carrasco and Johansson (BCJ) that the KLT relations are linked to a color-kinematics (CK) duality~\cite{Bern:2008qj}.
In a nutshell, Yang-Mills amplitudes can be organized as sums over trivalent graphs, dressed by color and kinematic (or BCJ) numerators that share common structural properties. 
The color numerators are built from group-theory tensors (such as structure constants) and therefore satisfy linear algebraic relations (such as the Jacobi identities).
The kinematic numerators are functions of the momenta and polarization vectors which can be chosen to fulfill exactly the same linear relations~\cite{Bern:2008qj}.
This implies in particular the aforementioned low-energy limit of the string monodromy relations, known as BCJ relations.
Finding BCJ numerators is typically non-trivial, but methods exist to derive them from known amplitudes~\cite{KiermaierTalk2010, Bjerrum-Bohr:2010pnr,Cachazo:2013iea,Naculich:2014rta}, or to construct them directly~\cite{Mafra:2011kj,Fu:2012uy,
Monteiro:2013rya,Fu:2017uzt,Du:2017kpo,
Chen:2019ywi,
Mizera:2019blq,Edison:2020ehu,He:2021lro,
Brandhuber:2021kpo,Chen:2021chy,Brandhuber:2021bsf}. 
The CK duality then permits to recover the KLT relations in an alternative way, namely through the replacement of color numerators by kinematic ones~\cite{Bern:2010ue,Bern:2010yg}.
The linear relations verified by all numerators then promote gauge invariance to diffeomorphism invariance.

After its discovery in YM amplitudes, a CK duality was shown to also be present in several other theories, including the non-linear sigma model (NLSM)~\cite{Chen:2013fya,Cheung:2016prv,Cheung:2017yef}, theories with matter particles~\cite{Johansson:2014zca,
Johansson:2015oia,
Chiodaroli:2015rdg,
delaCruz:2016wbr,
Brown:2016hck,
Brown:2018wss,
Johansson:2019dnu,
Plefka:2019wyg}, and the cubic theory of a bi-adjoint scalar (BAS)~\cite{Cachazo:2013iea} (see also~\cite{Bern:1999bx, Du:2011js, Bjerrum-Bohr:2012kaa}).
In the BAS theory, BCJ numerators are built out of group-theory structure constants only, and amplitudes generate the aforementioned KLT kernel.
Multiplying the numerators of two theories featuring a CK duality generates a whole web of double-copy theories, some of which are non-gravitational (see~\cite{Bern:2019prr} for a recent review).

The NLSM is a non-renormalizable theory, showing that the double copy applies to effective field theories (EFTs).
This is further confirmed by the terms of higher mass dimensions, i.e.\ of higher orders in $\alpha'$, in the low-energy expansion of the original KLT relations between open and closed string amplitudes.
Higher $\alpha'$ corrections appearing in the KLT kernel correspond to EFT operators in the cubic bi-adjoint theory~\cite{Mizera:2016jhj}.
This motivates the study of the double copy in EFTs.
The KLT formulation of the double copy was explored in this context and generalized in~\cite{Bjerrum-Bohr:2003hzh,Bjerrum-Bohr:2003utn,Chi:2021mio,Bonnefoy:2021qgu,Chen:2022shl,Chen:2023dcx}.
On the other hand, the CK duality has been studied for higher-derivative corrections to YM theory~\cite{Broedel:2012rc,He:2016iqi}, and bootstrap approaches towards gluon EFT numerators exist~\cite{Boels:2016xhc, Bern:2017tuc}.
More recently, the notion of CK duality was generalized by considering numerators which contain both kinematic and color information, including rules to build them for scalar particles~\cite{Carrasco:2019yyn,Low:2019wuv,Low:2020ubn,Carrasco:2021ptp,deNeeling:2022tsu,Li:2023wdm}.
For instance, these new numerators are needed for a CK-dual approach to a scalar EFT known as $Z$ theory, which plays a prominent role in the double copies of field theories to type I/II superstring theories, where it encodes all the necessary $\alpha'$ corrections~\cite{Broedel:2013tta,Carrasco:2016ldy,Mafra:2016mcc,Carrasco:2016ygv}.

EFTs are defined up to a cutoff scale $\Lambda$ above which a UV completion kicks in.
Calculations are then performed up to a fixed $(E/\Lambda)^n$ order, for some integer $n$ depending on the required precision and with $E$ the characteristic energy of a process.
As $n$ increases, EFT operators of higher dimensions can contribute and should be included.
In a bottom-up approach, agnostic about the underlying UV theory, the coefficients of different operators are all independent (and determined by measurements).
However, assumptions about the UV completion or, for example, on the soft behavior of the amplitudes~\cite{Cheung:2014dqa,Cheung:2015ota,Cheung:2016drk,Elvang:2018dco} typically correlate the operator coefficients.
Similarly, enforcing the CK duality also constrains the operators of a theory and their coefficients.
An infinite tower of higher-dimensional operators is for instance required for the tree-level double-copy consistency, to all EFT orders, of a YM theory including the dimension-six $F^3$ operator~\cite{Broedel:2012rc,Carrasco:2022lbm,Carrasco:2023wib}.
An elegant way to capture this tower is through the $(DF)^2$+YM theory~\cite{Johansson:2017srf}, which has been shown to supplement $Z$ theory in the double copies of field theories to bosonic and heterotic string theories~\cite{Azevedo:2018dgo}.
For the NLSM and other theories, the interplay between the double-copy consistency and soft behaviors has also been studied in~\cite{Elvang:2018dco,CarrilloGonzalez:2019fzc,Brown:2023srz}.

\begin{figure}[t]\centering
\begin{tikzpicture}
\node (d41) [ellipse, fill, text width=0mm]{};
\node (d42) [ellipse, fill, text width=0mm, right=of d41]{};
\node (d43) [ellipse, fill, text width=0mm, right=of d42]{};
\node (d61) [ellipse, fill, text width=0mm, below=of d41]{};
\node (d62) [ellipse, fill, text width=0mm, right=of d61]{};
\node (d81) [ellipse, fill, text width=0mm, below=of d61]{};
\node (t1) [above=2ex of d41, font=\LARGE]{1};
\node (t2) [above=2ex of d42, font=\LARGE]{2};
\node (t3) [above=2ex of d43, font=\LARGE]{3};
\node (d4) [left=2ex of d41, font=\LARGE]{4};
\node (d6) [left=2ex of d61, font=\LARGE]{6};
\node (d8) [left=2ex of d81, font=\LARGE]{8};
\draw [->] (t3) ++(5ex,0ex) -- node[below]{\minitab{flavor\\[-.75ex]traces}} ++(5ex,0ex);
\draw [->] (d8) ++(0ex,-5ex) -- node[right=.5ex]{\minitab{mass\\[-.75ex]dim.}} ++(0ex,-5ex);
\node (gbas) [above=1ex of t2, xshift=2ex, font=\LARGE]{GBAS};
\node [at=(d62), xshift=2ex, draw, inner sep=22ex, text width=0ex]{};
\node (y4) [ellipse, fill, text width=0mm, right=40ex of d41]{};
\node (y6) [ellipse, fill, text width=0mm, below=of y4]{};
\node (y8) [ellipse, fill, text width=0mm, below=of y6]{};
\node (y10) [ellipse, fill, text width=0mm, below=of y8]{};
\node [right=2ex of y4, font=\LARGE]{4};
\node [right=2ex of y6, font=\LARGE]{6};
\node [right=2ex of y8, font=\LARGE]{8};
\node [right=2ex of y10, font=\LARGE]{10};
\node [right=23.5ex of gbas,font=\LARGE]{YM};
\node [at=(y6), inner sep=0ex, xshift=1ex, text width=18ex, text height=44ex, draw]{};
\node (e1) [at=(d41), ellipse, red!60!black, draw, line width=.7mm, text width=2ex, inner sep=1.25ex, rotate=45]{};
\draw[->, dash pattern=on 8pt off 3pt, line width=.7mm, black!20!white] (e1) to[out=-30, in=180] (y4);
\node (e2) [at=(d41), yshift=-4.2ex, xshift=4.2ex, ellipse, blue!60!black, draw, line width=.7mm, text width=10ex, inner sep=1.75ex, rotate=45]{};
\node (empty) [below=20ex of d43, font=\LARGE, black!20!white]{$\emptyset$};
\draw[->, dash pattern=on 8pt off 3pt, line width=.7mm, black!20!white] (e2) to[out=-40, in=50] (empty);
\node (e3) [at=(d62), ellipse, green!60!black, draw, line width=.7mm, text width=17ex, inner sep=1.75ex, rotate=45]{};
\draw[->, dash pattern=on 8pt off 3pt, line width=.7mm, black!20!white] (e3) to[out=-50, in=130] (empty);
\draw[->, red!60!black, line width=.7mm] (e1) to[out=-30, in=180] (y6);
\draw[->, blue!60!black, line width=.7mm] (e2) to[out=-35, in=180] (y8);
\draw[->, green!60!black, line width=.7mm] (e3) to[out=-40, in=180] (y10);
\node[below=14ex of d43, xshift=1ex, rotate=-45, font=\LARGE]{$\cdots$};
\end{tikzpicture}
\caption{%
Illustration of the covariant-color-kinematics map from GBAS to YM tree-level EFT amplitudes, for towers of higher-dimensional operators satisfying the traditional color-kinematics duality.
The solid arrows represent the new flavor-kinematic replacement rule of \hyperref[trReplacementF^3]{Eqs.\,}\eqref{trReplacementF^3}, \eqref{trReplacementDim8}, and \eqref{trReplacementAllDim}.
It yields the all-multiplicity amplitude relations of \autoref{YMF3fromGBAS} for YM at dimension six (red arrow), of \autoref{YMF4fromGBAS} for YM at dimension eight (blue arrow), and the conjectured relation of \autoref{generalmRelationUnified} for the full EFT expansion of the $(DF)^2$+YM theory (green arrow and beyond).
The dashed arrows involve the replacement rule previously identified by Cheung and Mangan~\cite{Cheung:2021zvb} for one of the flavor traces.
Beyond the dimension-four single-trace order which they considered (upper dashed arrow), it leads to intricate cancellations between different GBAS amplitudes (lower dashed arrows), which originate in the relation displayed in \autoref{GBASF2fromGBAS} between dimension-six single-trace GBAS amplitudes and dimension-four double-trace ones. 
The latter are therefore sufficient to generate dimension-eight YM amplitudes, as expressed in \autoref{dim8fromdim4}.
This is also generalized to higher dimensions and up to six points in \autoref{generalmRelation2}.
}
\label{TowerSummary}
\end{figure}
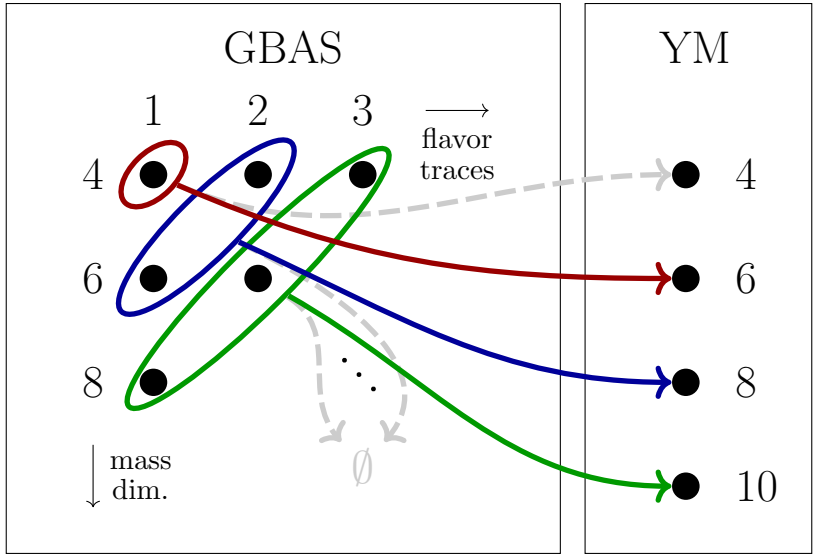

A different but closely related duality between color and kinematics was exposed by Cheung and Mangan~\cite{Cheung:2021zvb} at the level of classical equations of motion (EOMs) instead of amplitudes (see also~\cite{Mizera:2018jbh,Moynihan:2021rwh,Cao:2022vou,Wei:2023yfy} for related works).
Writing the NLSM EOM in terms of the chiral current and the YM EOM in terms of the field strength, they uncovered maps to the EOM of the BAS theory and to that of its gauged variant (the GBAS theory), respectively.
This so-called covariant color-kinematics (CCK) duality was further demonstrated to relate the color and kinematic algebras of the dual theories, as well as their conserved currents and, most importantly for the present work, their tree-level scattering amplitudes~\cite{Cheung:2021zvb}.
These can indeed be extracted from (the functional derivatives of) perturbative solutions to the EOMs with sources, and are therefore subject to the CCK duality.
New relations were derived between NLSM/BAS amplitudes as well as between YM/GBAS ones (upper dashed arrow in \autoref{TowerSummary}), through a replacement rule mapping flavor structures into kinematic ones.
Closed-form expressions for the BCJ numerators of the NLSM and YM theories at any multiplicity were also established.

In this paper, we study how the CCK duality extends to EFTs.\footnote{Following a different approach, the CK duality in off-shell currents of the YM EFT was previously studied in~\cite{Garozzo:2018uzj}.}
More precisely, we consider EFT corrections of increasing mass dimension to the YM and GBAS theories.
This analysis leads to a set of new relations between the tree-level amplitudes of the GBAS and YM EFTs.
These relations are summarized below.

In the YM EFT, the first higher-dimensional operator we consider involves three field-strength tensors, $F^3$.
We show that the dimension-six EOM it induces is mapped to the EOMs of the dimension-four GBAS theory, at the level of single flavor traces.%
\footnote{In this paper, we do not make use of the ordering with respect to the color indices shared by the scalar and the gluon.
Therefore, in what follows, ``traces" always implicitly refer to the flavor structures whose indices are only carried by the scalar.}
With this extended CCK duality at hand, we derive a relation between the tree-level amplitudes of the YM$+F^3$ and GBAS theories (solid red arrow in \autoref{TowerSummary}), which is realized by an additional replacement rule for flavor structures in terms of kinematic factors.
This result (as well as the ones below) have been confirmed by comparison against explicit Feynman diagram calculations of the relevant amplitudes.
Interestingly, the same single-trace dimension-four GBAS amplitudes which map to dimension-four YM amplitudes also encode dimension-six ones.

This mapping moreover allows us to derive a closed-form formula for the BCJ numerators of the YM+$F^3$ theory, at dimension six and any multiplicity.
These numerators are valid in $D$ space-time dimensions and are manifestly gauge invariant on all legs.

At dimension eight, we focus on the operators that satisfy the traditional CK duality.
We show that they lead to EOMs that are CCK-dual to those of a dimension-six extension of the GBAS theory, which is obtained by the dimensional reduction of the YM+$F^3$ theory from $D+n$ to $D$ dimensions.
The duality requires a correlated treatment of single and double traces of flavor structures in the GBAS theory.
Note that double-trace amplitudes were not involved in the CCK duality below dimension eight.
We also observe that the correlation between dimension-six and dimension-eight coefficients demanded to satisfy the regular CK duality are at the origin of cancellations that make the CCK duality possible. 
Leveraging this CCK duality, we obtain two different amplitude relations at any multiplicity.
One expresses dimension-eight YM amplitudes in terms of dimension-six single-trace and dimension-four double-trace GBAS amplitudes (solid blue arrow in \autoref{TowerSummary}), while the other relation requires only dimension-four double-trace GBAS amplitudes (after exploiting the cancellation illustrated by the lower dashed arrows in \autoref{TowerSummary}).
Remarkably, up to dimension eight, all the tree-level amplitudes of this YM EFT satisfying the regular CK duality are thus encoded in the renormalizable GBAS theory.
This also yields a straightforward procedure to derive dimension-eight BCJ numerators.

We conjecture a natural extension of the CCK duality beyond dimension eight, involving an increasing number of flavor traces.
In particular, explicit calculations up to six points show that the new amplitude relations discussed above extend to the full towers of higher-derivative operators defined by the $(DF)^2$+YM theory and its GBAS analog (solid green arrow and beyond in \autoref{TowerSummary}).
Various other explicit checks are performed.

The rest of this paper is organized as follows.
In \autoref{sec:review}, we first review the computation of tree-level scattering amplitudes from EOMs before turning to the CCK duality of Cheung and Mangan~\cite{Cheung:2021zvb}.
We then extend this duality to the EFT domain: \autoref{sec:F3} addresses YM$+F^3$ at dimension six, while the dimension-eight order is investigated in \autoref{sec:dim8}.
In \autoref{sec:beyonddim8}, we then study the CCK duality for the $(DF)^2$+YM theory.
We conclude in \autoref{sec:conclusions}.

\section{Review of the covariant color-kinematics duality}
\label{sec:review}

We start by reviewing how to solve EOMs perturbatively and how to extract tree-level scattering amplitudes from the resulting solutions \cite{Berends:1987me}.
After that, we review the CCK duality at the renormalizable level from~\cite{Cheung:2021zvb}.

\subsection{Tree-level scattering amplitudes from equations of motion}
\label{sec:amps-from-eom}

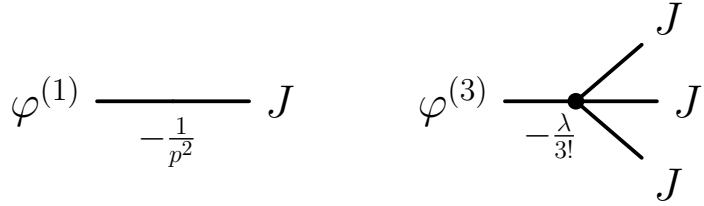
\begin{figure}
\centering\LARGE
\makebox[\textwidth][c]{
\fmfframe(5,5)(5,5){\begin{fmfgraph*}(20,15)
\fmfleft{l1}
\fmfright{r1}
\fmf{vanilla}{l1,v1,r1}
\fmfv{lab=$\varphi^{(1)}$}{l1}
\fmfv{lab=$J$}{r1}
\fmfv{lab=\scalebox{0.8}{$-\frac{1}{p^2}\hspace{2mm}$},lab.angl=-90}{v1}
\end{fmfgraph*}}
\hspace{2cm}
\fmfframe(5,5)(5,5){\begin{fmfgraph*}(20,15)
\fmfleft{l2}
\fmfright{r1,r2,r3}
\fmf{vanilla,tens=3}{l2,v1}
\fmf{vanilla}{v1,r1}
\fmf{vanilla}{v1,r2}
\fmf{vanilla}{v1,r3}
\fmfv{lab=$\varphi^{(3)}$}{l2}
\fmfv{lab=$J$}{r1}
\fmfv{lab=$J$}{r2}
\fmfv{lab=$J$}{r3}
\fmfv{d.shape=circle, d.size=1.7mm, d.fill=1, 
	lab=\scalebox{0.8}{$-\frac{\lambda}{3!}$\hspace*{-1mm}}, 
	lab.ang=-115,
    lab.dist=1.7mm}{v1}
\end{fmfgraph*}}
}
\caption{Diagrammatic representation of the perturbative solution to the EOM in the $\lambda\,\varphi^4$ theory.}
\label{fig:eom-diag}
\label{FeynmanDiagramsEOM}
\end{figure}

For simplicity, let us consider a massless real scalar field $\varphi$ with a quartic potential. 
The discussion below readily generalizes to other theories.
The corresponding Lagrangian reads
\beq
\cL=\frac{1}{2}\partial^\mu\varphi\partial_\mu\varphi-\frac{\lambda}{4!}\varphi^4+J\varphi \,,
\label{eq:phi4TheoryExample}
\eeq
from which the following EOM is obtained,
\beq
\Box\varphi+\frac{\lambda}{3!}\varphi^3=J \ . 
\label{toyScalarEom}
\eeq
The source $J(x)$ is non-dynamical and probes the response of the theory to an external perturbation.
At a given order $\cO(J^n)$ in the source, one can recursively compute the solution $\varphi^{(n)}$ to the EOM in perturbation theory:
\beq \label{SolveEOMPosition}
\bead
\varphi^{(1)}(x)&=\(\frac{1}{\Box}J\)\!(x)=-\int \text{d}^4y\frac{\text{d}^4p}{\(2\pi\)^4}\frac{e^{ip\cdot(x-y)}}{p^2}J(y) \ ,\\
\varphi^{(2)}(x)&=0 \ ,\\
\varphi^{(3)}(x)&=-\frac{\lambda}{3!}\(\frac{1}{\Box}\varphi^{(1)3}\)\!(x)=\frac{\lambda}{3!}\int \text{d}^4y\frac{\text{d}^4p}{\(2\pi\)^4}\frac{e^{ip\cdot(x-y)}}{p^2}\varphi^{(1)3}(y) \\
    &= -\frac{\lambda}{3!}\int 
    \left( \prod_{i=1}^3 \text{d}^4 y_i \frac{\text{d}^4p_i}{\(2\pi\)^4}\right) 
    \frac{1}{(p_1+p_2+p_3)^2} 
    \left(\prod_{i=1}^3 
        \frac{e^{ip_i(x-y_i)}}{p_i^2} J(y_i) \right) \ ,
\eead
\eeq
and so on.
In Fourier space, $\varphi(p)\equiv \int \text{d}^4x \, e^{-ip\cdot x}\varphi(x)$ and one finds
\beq
\bead
\varphi^{(1)}(p)&=-\frac{J(p)}{p^2} \ ,\\
\varphi^{(2)}(p)&=0 \ ,\\
\varphi^{(3)}(p)&=  
    -\frac{\lambda}{3!} 
    \int 
    \left( \prod_{i=1}^3\frac{\text{d}^4p_i}{\(2\pi\)^4}\right) 
    \frac{\delta^{(4)}(p-p_1-p_2-p_3)}{p^2} \frac{J(p_1)}{p_1^2}
    \frac{J(p_2)}{p_2^2}\frac{J(p_3)}{p_3^2}
\ ,
\\ & \ldots
\eead 
\eeq
These perturbative solutions can be represented in terms of Feynman graphs, as shown in 
\autoref{FeynmanDiagramsEOM}.
The tree-level scattering amplitudes of the theory are then obtained using the LSZ reduction formula.
At $n$ points and for all particles incoming,
\beq
\cA(p_1,...,p_n)=\int \prod_{i=1}^n \(\text{d}^4x_i \frac{ie^{-ip_i\cdot x_i} \, \Box_{x_i}}{(2\pi)^{3/2}}\)\langle 0|\text{T}\varphi(x_1)...\varphi(x_n)|0\rangle \ .
\eeq
The $n$-point correlator computed without source is obtained from the one-point function with a source, $\langle 0|\varphi(x_n)|0\rangle_J$, by taking functional derivatives:
\beq
\langle 0|\text{T}\varphi(x_1)...\varphi(x_n)|0\rangle=(-i)^{n-1}\(\frac{\delta^{n-1}}{\delta J(x_1)...\delta J(x_{n-1})}\langle 0|\varphi(x_n)|0\rangle_J\)\bigg\vert_{J=0} \ ,
\eeq
where, at tree-level, $\langle 0|\varphi(x_n)|0\rangle_J$ is simply the solution to the EOM with the source, evaluated at the point $x_n$.
Following the terminology of~\cite{Cheung:2021zvb}, we refer to $\varphi(x_n)$ as the \emph{root} leg of the corresponding diagrams, and to $\varphi(x_{1,...,n-1})$ as the \emph{leaf} legs.
For illustration,
\beq
\bead
\langle 0|\text{T}\varphi(x_1)...\varphi(x_4)|0\rangle&=(-i)^3\(\frac{\delta^{3}}{\delta J(x_1)...\delta J(x_3)}\langle 0|\varphi(x_4)|0\rangle_J\)\bigg\vert_{J=0}=(-i)^3\frac{\delta^{3}\varphi^{(3)}(x_4)}{\delta J(x_1)...\delta J(x_3)}\\
&=-i\lambda\int \prod_{i=1}^4
\(\frac{\text{d}^4p_ie^{-ip_i\cdot x_i}}{(2\pi)^4p_i^2}\) 
(2\pi)^4 \delta^{(4)}\!\!\(\sum_i p_i \) \ ,
\eead
\eeq
and
\beq
\bead
\cA(p_1,...,p_4)=\int \prod_{i=1}^4 \(\text{d}^4x_i \frac{-ie^{ip_i\cdot x_i}\;\Box_{x_i}}{(2\pi)^{3/2}}\)\langle 0|\text{T}\varphi(x_1)...\varphi(x_4)|0\rangle%\\
=-i\frac{\lambda}{(2\pi)^2}\;\delta^{(4)}\!\!\(\sum_ip_i\) \ ,
\eead
\eeq
consistently with the Feynman rules of the Lagrangian in \autoref{eq:phi4TheoryExample}.
In the rest of this paper, we write the amplitudes without momentum-conserving delta function and powers of $2\pi$ or $i$.

\begin{figure}
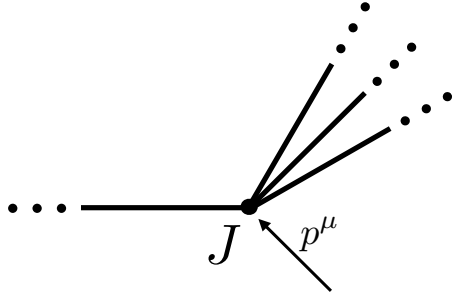

\centering
\adjincludegraphics[width=0.4\textwidth,trim={{.2\width} {.25\width} {.25\width} {.1\width}},clip]{figs/nonLinearJ.pdf}
\caption{Portion of diagram arising from a non-linear term $J\varphi^3$ involving the source $J$ in the equation of motion of $\varphi$.
It leads to diagrams proportional to $p^2$, where $p^\mu$ is the momentum flowing through the source.}
\label{nonLinearSourceDiag}
\end{figure}

Before closing this section, let us emphasize a point used later on: non-linear terms depending on the source in the EOMs are irrelevant on shell.
For concreteness, let us add the term $J\varphi$ to the right-hand side (r.h.s.)\ of the EOM in 
\autoref{toyScalarEom}.
This has the effect of turning on $\varphi^{(2)}$,
\beq
\varphi^{(2)}(x)=\frac{1}{\Box}\(J\varphi^{(1)}\)(x)=\int \text{d}^4y\frac{\text{d}^4p}{(2\pi)^4} \frac{e^{ip\cdot(x-y)}}{p^2}J(y)\int \text{d}^4z\frac{\text{d}^4q}{(2\pi)^4} \frac{e^{iq\cdot(y-z)}}{q^2}J(z) \ .
\eeq
Differentiating with respect to $J(x_2)$ and $J(x_3)$ and applying the LSZ formula, one finds
\beq
-i \, \cA(p_1,p_2,p_3)=p_2^2+p_3^2=0 \ ,
\eeq
i.e.\ the new term in the EOM has no effect on the on-shell scattering amplitudes.
More generally, terms of the form $J\varphi^n$ in the EOM would generate subdiagrams like that of \autoref{nonLinearSourceDiag}, which are proportional to the (vanishing) square of the momentum flowing through the source.

\subsection{Covariant color-kinematics duality between GBAS and YM}
\label{sec:originalCK}

We now turn to a review of the CCK duality introduced by Cheung and Mangan~\cite{Cheung:2021zvb}.
It establishes maps between EOMs of different theories and, therefore, between their tree-level scattering amplitudes.

Let us consider the Yang-Mills theory example, which will be most useful for our purposes.
Starting from the Yang-Mills (YM) Lagrangian with a source $J^a_\mu(x)$,
\beq
\cL_{\textsc{ym}}=
	-\frac{1}{4F^a_{\mu\nu}}F^{a\mu\nu}
	+A^{a\mu}J_\mu^a \ ,
\eeq
where the field-strength tensor is 
$F_{\mu\nu}^a\equiv 
    \partial_\mu A^a_\nu-\partial_\nu A^a_\mu+g\,f^{abc}A^b_\mu A^c_\nu$, 
one derives the usual YM EOM,
\beq \label{YMeom}
D^\mu F^{a}_{\mu\nu}=-J_\nu^a \ ,
\eeq
where 
$D_\mu F^a_{\nu\rho}\equiv 
    \partial_\mu F^a_{\nu\rho}+g\,f^{abc}A_\mu^bF^c_{\nu\rho}$ 
and $f^{abc}$ are group structure constants which verify the Jacobi identity.
Upon differentiating the EOM above and using the Bianchi identity, \cite{Cheung:2021zvb} showed that the following equation can be derived:
\beq \label{fieldstrengtheom}
D^2F^a_{\mu\nu}+g\,f^{abc}F^b_{\rho[\mu}F^{c\rho}_{\nu]}=-D_{[\mu}J_{\nu]}^a \ ,
\eeq
where we defined $X_{\[\mu\nu\]}\equiv X_{\mu\nu}-X_{\nu\mu}$, and where $DJ$ could be replaced by $\partial J$ without affecting the on-shell scattering amplitudes, as explained above.
This equation has the crucial property that the space-time indices of the gluon field strength are not contracted with those of covariant derivatives.
Since $D^2=\Box+$non-linear interaction terms dependent on $A_\mu$ and $F_{\mu\nu}$, given a solution $A_\mu$ and $F_{\mu\nu}$ at a given order in the source, one can solve for $F_{\mu\nu}$ at the next order by simply inverting $\Box$, without making the relation between $F_{\mu\nu}$ and $A_\mu$ explicit.
Consequently, one can reinterpret \autoref{fieldstrengtheom} as describing the propagation of six flavors of colored scalars 
\begin{equation}
\lambda\,\phi^{aA}
\leftrightarrow 
F^a_{\mu\nu}
\,,
\end{equation}
with a cubic interaction.
(We have included a factor of $\lambda$ in accordance with dimensional analysis.)
Moreover, that cubic interaction can be expressed in terms of a structure constant $f^{ABC}$, to be constructed below, which verifies the Jacobi identity. 
Therefore, the scalars form a bi-adjoint multiplet whose first symmetry group has been gauged.
Below, we refer to those two groups as color and flavor, respectively.
This theory is known as the gauged bi-adjoint scalar theory (GBAS).
Its Lagrangian reads
\beq
\cL_\textsc{gbas}=\cL_{\textsc{ym}}+\frac{1}{2}D^\mu\phi^{aA}D_\mu\phi^{aA}-\frac{g\, \lambda}{3}f^{abc}f^{ABC}\phi^{aA}\phi^{bB}\phi^{cC}+J^{aA}\phi^{aA} 
\eeq
where $\lambda$ has mass dimension one and leads to the following EOM
\begin{align}
D^2\phi^{aA}+g\, \lambda \, f^{abc}f^{ABC}\phi^{bB}\phi^{cC}
&=J^{aA} \ ,
\label{eq:GBASeomDim4}
\end{align}
from which we can read off the map of the scalar source into the gluon one,
\beq
\lambda\, J^{aA}
\leftrightarrow
-D_{[\mu}J^a_{\nu]} \ ,
\label{eq:sourceRelationGluons}
\eeq
as well as the map for the flavor structure constant in terms of the space-time metric,
\beq
f^{A_1A_2A_3}
\leftrightarrow
-\frac{1}{4}\, \eta^{\nu_3][\mu_1}\eta^{\nu_1][\mu_2}\eta^{\nu_2][\mu_3}
\ .
\label{eq:structureConstantRelation}
\eeq
Having connected the EOMs of the two theories, we can also connect their one-point functions with sources, and therefore their scattering amplitudes.
This is however nontrivial given {\it i}\,) that the bi-adjoint scalar still interacts with gluons, and {\it ii}\,) that the sources for both fields are correlated according to \autoref{eq:sourceRelationGluons}. 

The complication {\it i}\,) arises since we artificially separated the gluon field and its field strength.
In order to compute scattering amplitudes as sketched in \autoref{sec:amps-from-eom}, we could use 
$\langle0|A_\mu^a|0\rangle_J$ or 
$\langle0|F_{\mu\nu}^a|0\rangle_J$.
Both fields interpolate single-gluon states and can be related after gauge fixing.
So using either of them simply changes the differential operators that act on the $n$th field in the LSZ reduction formula.
For instance, in an axial gauge where $q^\mu A_\mu^a=0$ for an arbitrary reference vector $q$,
\beq
|g^a(p,h)\rangle=\epsilon^\nu_h A^a_\nu(p)|0\rangle=\frac{i q^{[\mu}_{\phantom{h}}\epsilon^{\nu]}_h F^a_{\mu\nu}(p)}{2\, q\cdot p}|0\rangle
\ ,
\eeq
for a gluon of momentum $p$, helicity $h$ and color $a$. 
Reference~\cite{Cheung:2021zvb} proposes to use the field strength, related to $\langle0|\phi^{aA}|0\rangle_J$ in the dual theory through the CCK replacement rule,
\beq
\lambda\,\Big[
	   \langle0|\phi^{aA}|0\rangle_J
\Big]_\text{GBAS}
\quad
\tikz[baseline]\draw[->, dash pattern=on 8pt off 3pt, line width=.7mm] (0,.5ex)-- node[black, below=0mm, font=\scriptsize]{CCK} (1.5cm,.5ex);
\quad
\Big[
	\langle0|F_{\mu\nu}^a|0\rangle_J
\Big]_\text{YM}
\,,
\label{dim4OnePtFctEq}
\eeq
which we make explicit below.
Differentiating with respect to sources, this implies a duality between GBAS scattering amplitudes involving at least one scalar and YM amplitudes.
Importantly, one should note that the computation of $\langle0|\phi^{aA}|0\rangle_J$ is affected by the fact {\it ii}\,): in the perturbative solution for $\phi^{aA}$, the same source generates both gluons and scalars.
Therefore, $n$-point scattering amplitudes of gluons in the YM theory are mapped to combinations of amplitudes with different numbers of scalars in the GBAS theory; specifically $2\leq m \leq n$ scalars and $n-m$ gluons (where we used the fact that tree-level GBAS amplitudes with a single scalar are zero).

Although the GBAS scalar EOM of \autoref{eq:GBASeomDim4} is in one-to-one correspondence with the YM field strength EOM of \autoref{fieldstrengtheom}, the gluon EOMs in the two theories are different.
The YM gluon propagates according to \autoref{YMeom}, whereas the GBAS gluon EOM in principle includes a scalar current of the form $\phi D\phi$.
However, this term can be ignored when restricting to single-trace GBAS amplitudes calculated from $\langle0|\phi^{aA}|0\rangle_J$, in which case the two gluon EOMs effectively become identical.
The CCK duality can thus be phrased as a map from GBAS amplitudes with only a single trace of flavor group generators to YM amplitudes.

What happens in practice is best described through an example, so let us focus on the three-point gluon amplitude $\cA_\textsc{ym}(1,2,3)$.
It can be computed from the three-point correlator $\langle 0|T A^a_\mu(x) A^b_\nu(y) F^c_{\rho\sigma}(z)|0\rangle$, using the usual LSZ reduction formula with the exception that the third polarization should be replaced by $iq^{[\rho}_{\phantom{3}}\epsilon^{\sigma]}_3/(2\,q\cdot p_3)$.
That correlator can itself be derived from $\langle 0|F^c_{\rho\sigma}(z)|0\rangle_J$, upon differentiation with respect to $J^a_\mu(x)$ and $J^b_\nu(y)$, before fixing all sources to zero.
By the CCK duality of the EOMs, this is equivalent to acting on $\langle 0|\phi^{cC}|0\rangle_J$.
Now, which amplitudes of the regular gauged bi-adjoint theory are generated by $\langle 0|\phi^{cC}|0\rangle_J$? We have that
\begin{multline}
(-i)\frac{\delta}{\delta J^{b\nu}(y)}\langle 0|\phi^{cC}(z)|0\rangle_J
=\\\int \text{d}^4y'\[\frac{\delta J^{b'\nu'}(y')}{\delta J^{b\nu}(y)}\langle 0|A^{b'}_{\nu'}(y')\phi^{cC}(z)|0\rangle_J+\frac{\delta J^{b'B}(y')}{\delta J^{b\nu}(y)}\langle 0|\phi^{b'B}(y')\phi^{cC}(z)|0\rangle_J\]
\end{multline}
where, ignoring non-linear terms involving the source,
\beq
\frac{\delta J^{b'\nu'}(y')}{\delta J^{b\nu}(y)}=\delta^{b'}_b\delta^{\nu'}_{\nu}\delta^{(4)}(y-y') \ , \quad \frac{\delta J^{b'B}(y')}{\delta J^{b\nu}(y)}
=-\delta^{b'}_b\delta^{B,\rho\sigma}\partial_{[\rho}\eta_{\sigma]\nu}\delta^{(4)}(y-y')\,.
\eeq
The second equation here arises from the relation between the sources in \autoref{eq:sourceRelationGluons} and results in an external polarization of the scalars given by $-ip_{[\mu}\epsilon_{\nu]}$.
Differentiating once more, using the LSZ formula and matching to GBAS amplitudes, 
one finds
\beq
\bead
\cA_{\textsc{ym}}(g_1,g_2,g_3)=&
\frac{i\delta^{\alpha\beta}_{A_3}q_{[\alpha}\epsilon_{3\beta]}}{2\,q\cdot p_3}
\Bigg[\delta^{\mu\nu}_{A_1}\delta^{\rho\sigma}_{A_2}\(-ip_{1[\mu}\epsilon_{\nu]1}\)\(-ip_{2[\rho}\epsilon_{\sigma]2}\)
\cA_{\textsc{gbas}}(\phi_1^{A_1},\phi_2^{A_2},\phi_3^{A_3})
\\&+\Big\{\delta^{\rho\sigma}_{A_2}\(-ip_{2[\rho}\epsilon_{\sigma]2}\)
	\cA_{\textsc{gbas}}(g_1,\phi_2^{A_2},\phi_3^{A_3})
+(1\leftrightarrow 2)\Big\}
\\&+\cA_{\textsc{gbas}}(g_1,g_2,\phi_3^{A_3})\Bigg] \ ,
\eead
\label{eq:YMfromGBAS1}
\eeq
where the last amplitude on the r.h.s.\ actually vanishes.

In general, an explicit restriction to single traces has to be performed on the GBAS side.
However, the amplitudes on the r.h.s.\ of \autoref{eq:YMfromGBAS1} only involve a single trace of flavor generators and can therefore be kept.
Actually, one obtains simpler formulae by making those flavor factors explicit, i.e.\ using flavor-ordered GBAS amplitudes.
Let us look at the first line of \autoref{eq:YMfromGBAS1} above: $\cA(\phi_1^{A_1},\phi_2^{A_2},\phi_3^{A_3})$ comes with a factor of $\lambda\,f^{A_1A_2A_3}$ defined in \autoref{eq:structureConstantRelation}.
Contracting with the momentum and polarization factors, one finds $-4i\,\Tr(\F_1\F_2\tilde \F_3)$ where
\beq
\F_i^{\mu\nu}\equiv p_i^{[\mu}\epsilon_i^{\nu]} \ , 
\quad \tilde \F_i^{\mu\nu}\equiv -\frac{q^{[\mu}_{\phantom{i}}\epsilon_i^{\nu]}}{2\, p_i\cdot q} \ , 
\quad \Tr(\cO_{\mu\nu})\equiv \frac{1}{2}\eta^{\mu\nu}\cO_{\mu\nu} \ .
\eeq
Using the antisymmetry of $\ptwiddle{\F}$, one can rewrite $-4i\,\Tr(\F_1\F_2\tilde \F_3)=-2i\,\Tr(\F_1\F_2\tilde \F_3-\F_2\F_1\tilde \F_3)$ and compare with the usual relation, $f^{A_1A_2A_3}=-2i\,\Tr(T^{A_1}T^{A_2}T^{A_3}-T^{A_2}T^{A_1}T^{A_3})$.
The flavor factor of the term on the second line of \autoref{eq:YMfromGBAS1} is $\delta^{A_2A_3}=-\frac{1}{4}\eta^{\nu_3][\mu_2}\eta^{\nu_2][\mu_3}$, so that we find $2\,\Tr(\F_2\tilde \F_3)$, to be compared with $\delta^{A_2A_3}=2\,\Tr(T^{A_2}T^{A_3})$.
We can therefore rewrite \autoref{eq:YMfromGBAS1} as 
\beq
\cA_{\textsc{gbas}}(\phi_1^{A_1},\phi_2^{A_2},\phi_3^{A_3})+\Big\{\cA_{\textsc{gbas}}(g_1,\phi_2^{A_2},\phi_3^{A_3})+(1\leftrightarrow 2)\Big\}
\quad
\tikz[baseline]\draw[->, dash pattern=on 8pt off 3pt, line width=.7mm] (0,.5ex)-- node[black, below=0mm, font=\scriptsize]{CCK} (1.5cm,.5ex);
\quad
\cA_{\textsc{ym}}(g_1,g_2,g_3)
\ ,
\label{eq:YMfromGBAS2}
\eeq
where the CCK replacement rule on the flavor factors is
\beq \label{trReplacement}
\lambda^{n-2} \, \text{Tr}(T^{A_1}T^{A_2}...T^{A_n}) 
%\to
\quad
\tikz[baseline]\draw[->, dash pattern=on 8pt off 3pt, line width=.7mm] (0,.5ex)-- node[black, below=0mm, font=\scriptsize]{CCK} (1.5cm,.5ex);
\quad
\text{Tr}(\F_1 \F_2...\tilde \F_n)
\, .
\eeq 
In terms of flavor-ordered amplitudes, this means 
\begin{align}
 \cA_3^\textsc{ym}
		=  A_{\phi\phi\phi}[123]\,\trFt{12\tilde 3}
			+ A_{\phi\phi\phi}[213]\,\trFt{21\tilde 3} 
   +A_{\phi g\phi}[13]\,\trFt{1\tilde 3}
			+A_{g\phi\phi}[23]\,\trFt{2\tilde3}\,,
\label{eq:dim4TraceReplacement}
\end{align}
where we used the shorthand notation $\trFt{\sigma \tilde n} \equiv \text{Tr}(\F_{\sigma_1}...\F_{\sigma_{|\sigma|}}\tilde \F_n)$ with $\sigma$ being the permutation of the $\phi$-scalar subset of the $(1,...,n-1)$ particles.
The flavor-ordered GBAS amplitudes on the r.h.s.\ have $[\sigma n]$ arguments specifying the flavor traces that have been isolated (along with powers of $\lambda$), and have subscript making explicit which of the particles are scalars and gluons.

Diagrammatically, the CCK map for this three-point amplitude (or one-point function expanded to 
$\cO(J^2)$) is thus the following:
\begin{equation}\label{CCKdim4Diagramatically}
\begin{tikzpicture}[baseline]
\node (l1) {\begin{fmfgraph*}(20,15)
	\fmfstraight
	\fmfleft{l1}
	\fmfright{r1,r2}
	\fmf{dashes,tens=2}{l1,v1}
	\fmf{dashes}{v1,r1}
	\fmf{dashes}{v1,r2}
	\fmfv{d.shape=circle, d.size=1.7mm, fore=black,
		lab=$\phi\phi$, lab.angl=-110, lab.dist=3mm
		}{v1}
	\end{fmfgraph*}};
\node (l2) [right=2mm of l1]{$+$};
\node (l3) [right=2mm of l2] {\begin{fmfgraph*}(20,15)
	\fmfstraight
	\fmfleft{l1}
	\fmfright{r1,r2}
	\fmf{dashes,tens=2}{l1,v1}
	\fmf{dashes}{v1,r1}
	\fmf{photon}{v1,r2}
	\fmfv{d.shape=circle, d.size=1.7mm, fore=black,
		lab=$D^2\phi$, lab.angl=-110, lab.dist=3mm
		}{v1}
	\end{fmfgraph*}};
\node (r1)[right=2.5cm of l3]{\begin{fmfgraph*}(20,15)
	\fmfstraight
	\fmfleft{l1}
	\fmfright{r1,r2}
	\fmf{plain,tens=2}{l1,v1}
	\fmf{plain}{v1,r1}
	\fmf{plain}{v1,r2}
	\fmfv{d.shape=circle, d.size=1.7mm, fore=black,
		lab=$F F$, lab.angl=-110, lab.dist=3mm
		}{v1}
	\end{fmfgraph*}};
\node (r2) [right=2mm of r1]{$+$};
\node (r3) [right=2mm of r2]{\begin{fmfgraph*}(20,15)
	\fmfstraight
	\fmfleft{l1}
	\fmfright{r1,r2}
	\fmf{plain,tens=2}{l1,v1}
	\fmf{plain}{v1,r1}
	\fmf{photon}{v1,r2}
	\fmfv{d.shape=circle, d.size=1.7mm, fore=black,
		lab=$D^2 F$, lab.angl=-110, lab.dist=3mm
		}{v1}
	\end{fmfgraph*}};
\draw[->, dash pattern=on 8pt off 3pt, line width=.7mm] ($(l3.east)+(5mm,0mm)$) -- node[black, below=0mm, font=\scriptsize]{CCK} ($(r1.west)+(-5mm,0mm)$);
\node [above=9mm of l2.center]{GBAS};
\node [above=9mm of r2.center]{YM};
\node [right=2mm of r3]{,};
\end{tikzpicture}
\end{equation}
where the vertices are schematically labeled by the terms which generate them in the YM-field-strength and GBAS EOMs of \autoref{fieldstrengtheom} and \autoref{eq:GBASeomDim4}.
Dashed lines represent $\phi$ scalars, solid ones represent the YM field strength $F$, while wavy ones are gluons.
Note that on the r.h.s.\ only pure-gluon amplitudes are generated (external gluons are interpolated by \emph{both} $A_\mu$ and $F_{\mu\nu}$), while on the left-hand side (l.h.s.) we start from pure-scalar and mixed scalar-gluon amplitudes.
The EOM evolution of the fields is pictured in these diagrams from left to right, from the initial root leg to the final leaf legs (or sources).
Starting from a scalar root leg, the restriction to the single-trace sector of the GBAS theory is achieved by allowing scalar legs to branch into scalars and gluons, while forbidding gluons to branch back into scalars.

Extending the above to $n$-point scatterings, \cite{Cheung:2021zvb} found
\beq \label{YMampfromGBAS}
	A_{\textsc{ym},n}
 = 
	\sum_{\Phi\in \mathbb{P}^+(1...n-1)}
	\;
	\sum_{\sigma\in S(\Phi)}
	A_\textsc{gbas}[\sigma n] \, \trFt{\sigma\tilde n}
 \,,
\eeq 
where the first sum runs over all different choices of $m-1$ scalars (with $2 \leq m \leq n$), captured by the non-empty power set $\mathbb{P}^+(1...n{-}1)$, which is the set of all non-empty subsets of $(1,...,n{-}1)$, while the $S(\Phi)$ set captures all permutations of $\Phi$.
We stress once more that factors of the dimensionful coupling $\lambda$ are taken out of the above formula through the definition of the flavor-ordered amplitude, as required for a correct matching of amplitude dimensions.
The same applies to all formulae of that sort in what follows.

\subsection{Derivation of Yang-Mills numerators}
\label{s:numDim4}

The CCK duality as presented in \autoref{YMampfromGBAS} derives YM amplitudes from GBAS ones with fewer gluons and more scalars, but not quite from pure-scalar amplitudes yet.
Conversely, it is also known how to relate amplitudes in the opposite direction: namely to obtain GBAS amplitudes with fewer gluons and more scalars, or to get GBAS amplitudes from YM ones, through the so-called \emph{transmutation} operators of Ref.~\cite{Cheung:2017ems}.
Combining both techniques, Cheung and Mangan~\cite{Cheung:2021zvb} derived a closed-form expression for the BCJ numerators of YM at any multiplicity in the trace basis.
These allow for an explicit decomposition of gluon amplitudes in terms of single-trace pure-scalar GBAS (i.e.\ BAS) amplitudes,
\beq 
\mathcal{A}_{\textsc{ym},n} = 
\sum_{\sigma \in S(1...n-1)}
A^\textsc{bas}_{\phi^n}[\sigma n] \,
K^{(4)}[\sigma n] 
\,.
\eeq
The numerator superscript $(4)$ distinguishes it from analogous objects derived below at higher EFT order.
From these trace-basis numerators $K^{(4)}[\sigma n]$ with any ordering $\sigma$, one can straightforwardly obtain BCJ numerators for the YM theory (in the adjoint basis).%
\footnote{The procedure is identical to the one through which one generates adjoint color structures from traces of color generators~\cite{Bern:2011ia, Naculich:2014rta, Bonnefoy:2021qgu}.
In this analogy, $K^{(4)}$ plays the role of a trace and the resulting BCJ numerators have the required adjoint-like properties.
Note that the trace-basis numerators $K^{(4)}$ are more redundant than regular BCJ ones, since certain trace-like structures give rise to vanishing adjoint-like objects.}
They are therefore directly relevant for the regular CK duality and the BCJ approach to the double copy.

Because we will follow the same procedure to derive numerators in the EFT below, we now review this at three points.
Let us consider the expression of the three-point YM amplitude in terms of the polarization vector $\epsilon_i$ of the gluon $i$.
It has been shown in \cite{Cheung:2017ems} that acting with the operator $\partial_{\epsilon_1\cdot \epsilon_3}$ on that amplitude generates a GBAS amplitude according to the transmutation relation 
$-2\,\partial_{\epsilon_1\cdot \epsilon_3} 
\mathcal{A}_{\textsc{ym},3} = A^\textsc{gbas}_{\phi g\phi}[13]$.
Acting now with this operator on both sides of \autoref{eq:dim4TraceReplacement}, we can solve for the mixed scalar-gluon amplitude in terms of a pure-scalar amplitude,
\beq 
\label{GBAStoBAS-3pt}
A^\textsc{gbas}_{\phi g\phi}[13]
= 
A^\textsc{bas}_{\phi^3}[123]
\:\G[1,2,3]\,,
\eeq 
where
    \begin{align}\label{Gdef}
        \G[\sigma, \tau, \rho] 
        \equiv 
        - \frac{(p_{\sigma})_{\mu} (\F_{\tau})^{\mu\nu} \, q_\nu}
               {(p_{\sigma\rho})^\alpha \, q_\alpha}
        \,,
    \end{align}
with $q$ an arbitrary reference momentum, $p_\sigma = p_{\sigma_1}+...+p_{\sigma_{|\sigma|}}$
and 
\beq 
(\F_{\sigma})^{\mu\nu} = 
(\F_{\sigma_1})^{\mu}_{\ \mu_1}
    (\F_{\sigma_2})^{\mu_1}_{\ \mu_2}
    ...(\F_{\sigma_{|\sigma|}})^{\mu_{|\sigma|-1}\nu}\,.
\eeq 
Inserting this back into \autoref{eq:dim4TraceReplacement}, we obtain the YM amplitude in terms of pure-scalar GBAS amplitudes and, hence, the three-point numerator:
\beq 
K^{(4)}[123] = \trFt{12 \tilde 3}
+ \G[1,2,3]\: \trFt{1\tilde 3}
\,.
\eeq 
The other numerator, $K^{(4)}[213]$, can be derived in a similar way, or it can simply be obtained as a permutation of the particle labels in the above numerator.

\section{Effective-field-theory extension to dimension six}
\label{sec:F3}

The derivation above relies on the precise form of the EOMs, i.e.\ of the interactions.
It is therefore natural to ask whether these can be modified while maintaining the CCK duality.
One possible modification is to deform the action by the addition of higher-dimensional operators, while keeping the spectrum untouched.
It is known that a regular CK duality exists at least for some of those deformations, including the lowest-order dimension-six correction to the Yang-Mills theory consisting of a trace of three field-strength tensors~\cite{Broedel:2012rc}.

In this section, we thus consider the $\cO(1/\Lambda^2)$ amplitudes of such a YM$+F^3$ theory:%
\footnote{To avoid confusion with the terminology used there, we stress that the $F^3$ operator is not related to the $F^3$ replacement rule of~\cite{Cheung:2021zvb}, where higher-derivative interactions are not considered.}
\beq 
\label{eq:F3Lag}
	\mathcal{L}_{\textsc{ym}}^{(6)} = 
			-\frac{1}{4}F_{\mu\nu}^aF^{a\mu\nu}
			- \frac{g%\,c_6
   }{3\, \Lambda^2} 
				f^{abc}F^{a\ \nu}_{\ \mu}
				F^{b\ \rho}_{\ \nu} F^{c\ \mu}_{\ \rho}
			+ A^{a}_\mu J^{a \, \mu}_{A}
				\,.
\eeq
where $\Lambda$ is an energy scale.
We will find that a CCK duality is still present, which will be expressed in terms of scattering amplitudes at the end of \autoref{sec:dim6Details}.
In terms of one-point functions in the presence of sources, it reads
\beq
\Big[
	   \langle0|A_\mu^a|0\rangle_J^{(4, 1)}
\Big]_\text{GBAS}
\quad
\tikz[baseline]\draw[->, line width=.7mm] (0,.5ex)-- node[below=0mm, font=\scriptsize]{CCK} (1.5cm,.5ex);
\quad
\Big[
	\langle0|A_\mu^a|0\rangle_J^{(6)}
\Big]_\text{YM}
\,,
\label{dim6OnePtFctEq}
\eeq
where $\big[\langle 0|\chi|0\rangle_J^{(m[,n])}\big]_\text{Th.}$ denotes the one-point function of the field $\chi$ computed in the theory Th.\ at mass dimension $m$, in the $n$-trace sector (only for the GBAS theory).
This duality therefore relates the \emph{renormalizable} GBAS to the dimension-six YM+$F^3$ effective field theory.
The subscript $J$ indicates that the one-point function is computed in the presence of sources and, in the GBAS theory, the two sources are correlated as in \autoref{eq:sourceRelationGluons}.
Finally, the CCK map is extended to a new treatment of flavor traces, different from that of \autoref{eq:dim4TraceReplacement}, which is presented below.

\subsection{Covariant color-kinematics duality between GBAS and YM\texorpdfstring{$+F^3$}{+FFF}}\label{sec:dim6Details}
 
To establish this CCK duality, we inspect the EOM of the YM$+F^3$ theory,
\beq 
	D^\mu F_{\mu\nu}^a
	+\frac{g%c_6
 }{\Lambda^2}\,
	f^{abc}\,F^b_{\mu\rho}
	D_\nu F^{c,\mu\rho}  
	% = -(J_A)^a_\nu
	= -J^a_\nu
\ ,
\label{F3eom} 
\eeq 
derived using the Bianchi identity, dropping non-linear terms where sources multiply other fields, and truncating to $\cO(1/\Lambda^2)$ by using the renormalizable YM EOM of \autoref{YMeom} in terms that are already suppressed by $1/\Lambda^2$.

As field-strength and covariant-derivative indices are not contracted together in the second term of the l.h.s., this EOM can be mapped to the gluon EOM in the GBAS theory,
\beq \label{eomGBASgluon}
	D^\mu F_{\mu\nu}^a
	+g\,f^{abc}\,\phi^{bA}D_\nu\phi^{cA}
	= -J^a_\nu
\,,
\eeq 
through the same replacement as in the previous section, $F^a_{\mu\nu} \leftrightarrow \lambda\,\phi^{aA}$, but with the notable difference that the field strength in the first term, $D^\mu F_{\mu\nu}^a$, does not get mapped.
Instead, the variable of interest in this term remains the gluon field $A_\mu$ and not $F_{\mu\nu}$. 
The EFT power counting makes this partial map consistent, when solving the EOM perturbatively in $J$ {\it and} in $1/\Lambda^2$ as follows.
Denoting $F^{(d)}$ the solution at order $\cO\!\(1/\Lambda^{d-4}\)$, the EOM of \autoref{F3eom} can be rewritten as
\beq 
\bead
	&D^\mu F^{(6)a}_{\mu\nu}
	+\frac{g%c_6
 }{\Lambda^2}
	\,f^{abc}\,F^{(4)b}_{\alpha\beta}
	D_\nu F^{(4)c,\alpha\beta} 
	= 0
\,, \\
	&D^\mu F^{(4)a}_{\mu\nu}
	= -J^a_\nu\,.
\label{F3eomPowerCounting} 
\eead
\eeq 
We do not consider $F^{(d>6)}$ since we have dropped terms of order $\cO\!\(1/\Lambda^{4}\)$ when deriving \autoref{F3eom}.
Following the steps of the previous section, we can therefore interpret $F^{(4)}$ as a scalar propagating in a gluon background, while considering $F^{(6)}$ as the field-strength tensor of that gluon.%
\footnote{One may wonder why we do not also try to interpret $F^{(6)}$ as a scalar.
It turns out that manipulating the dimension-six gluon EOM of \autoref{F3eom} as done to obtain \autoref{fieldstrengtheom} leads to the following EOM for $F^{(6)}$,
\bes
D^2F^a_{\mu\nu}+f^{abc}F^b_{\rho[\mu}F^{c\rho}_{\nu]}
-\frac{g%c_6
}{\Lambda^2}f^{abc}\(f^{bde}F_{\mu\nu}^dF^{e}_{\rho\sigma} F^{c\rho\sigma}
+D_{[\mu}F^{c\rho\sigma}D_{\nu]}F^{b}_{\rho\sigma}\)=-D_{[\mu}J_{\nu]}^a \ ,
\ees
which cannot easily be recast as a scalar EOM because of the presence of covariant derivatives with uncontracted indices.}
The EOM of $F^{(4)}$ can then be rewritten as
\beq
D^2F^{(4)a}_{\mu\nu}+g\,f^{abc}F^{(4)b}_{\rho[\mu}F^{(4)c\rho}_{\nu]}=-D_{[\mu}J_{\nu]}^a
\ ,
\eeq
just as in the renormalizable case discussed in \autoref{sec:originalCK}.

Thanks to this duality between the gluon EOM in YM$+F^3$ and the gluon EOM in GBAS (\autoref{F3eom} and \autoref{eomGBASgluon}), YM$+F^3$ amplitudes are therefore encoded in GBAS ones.
To be precise and as anticipated in \autoref{dim6OnePtFctEq}, the EFT power counting implies that the relevant GBAS amplitudes are those obtained from the single-trace part of $\langle0|A_\mu^a|0\rangle_J$ with at least two scalars.
This means that $n$-gluon amplitudes in YM$+F^3$ are mapped to combinations of amplitudes with $2\leq m \leq n-1$ scalars and $n-m\geq 1$ gluon(s). 
We stress here the difference with \autoref{sec:originalCK}, where the relevant GBAS object is the single-trace part of the scalar one-point function $\langle 0 | \phi^{aA} |0\rangle_J$, and where the relevant amplitudes have $2 \leq m \leq n$ scalars and $n-m\geq 0$ gluons.

To relate the GBAS amplitudes to pure-gluon ones, the external polarizations are again determined by \autoref{eq:sourceRelationGluons} in the same way as in the previous section.
Similarly to \autoref{trReplacement}, the flavor traces are replaced by combinations of momenta and polarization vectors,
\beq \label{trReplacementF^3}
\lambda^{n-2} \, \text{Tr}(T^{A_1}T^{A_2}...T^{A_n})
\quad
\tikz[baseline]\draw[->, line width=.7mm] (0,.5ex)-- node[below=0mm, font=\scriptsize]{CCK} (1.5cm,.5ex);
\quad
\frac{1}{\Lambda^2} \, \text{Tr}(\F_1 \F_2...\F_n)
\, ,
\eeq
but $\tilde \F$ no longer appears since the generating correlator $\langle0|A_\mu^a|0\rangle_J$ now features the gluon field.
The explicit factor of $1/\Lambda^2$ clearly shows that this CCK duality generates higher-derivative interactions.

It thus follows that the $n$-point dimension-six YM$+F^3$ amplitude is encoded in single-trace GBAS amplitudes through
\beq \label{YMF3fromGBAS}
	\mathcal{A}_{\textsc{ym},n}^{(6)} = 
	\frac{1}{\Lambda^2}
	\sum_{\Phi\in \mathbb{P}^{++}(1...n-1)}
	\;
	\sum_{\sigma\in S(\Phi)/Z_{|\Phi|}}
	A_\textsc{gbas}[\sigma]
	\;
	\F[\sigma]
\,.
\eeq 
where $ \F[\sigma] \equiv \text{Tr}(\F_{\sigma_1}...\F_{\sigma_{|\sigma|}})$.
This equation is similar to \autoref{YMampfromGBAS}, with important differences arising from the fact that the $n$th particle is now a gluon. (Note that no new GBAS amplitude is needed beyond those appearing in \autoref{YMampfromGBAS}, particle relabeling is sufficient.)
Since at least two scalars are required in the GBAS amplitudes, there appears the set of all subsets of $(1...n{-}1)$ containing at least two elements, denoted $\mathbb{P}^{++}(1,...,n{-}1)$.
In addition, the set $S(\Phi)/Z_{|\Phi|}$ contains all permutations that result in inequivalent traces (using cyclicity).
The three-point CCK map at dimension-six is for instance the following: 
\begin{equation}
\begin{tikzpicture}[baseline]
\node (l3) [right=2mm of l2]{\begin{fmfgraph*}(20,15)%
	\fmfstraight
	\fmfleft{l1}
	\fmfright{r1,r2}
	\fmf{photon,tens=2}{l1,v1}
	\fmf{dashes}{v1,r1}
	\fmf{dashes}{v1,r2}
	\fmfv{d.shape=circle, d.size=1.7mm, fore=black,
		lab=$\phi D\phi$\hspace{-3.3mm}, lab.angl=-115, lab.dist=3mm
		}{v1}
	\end{fmfgraph*}};
\node (r1) [right=2.5cm of l3]{\begin{fmfgraph*}(20,15)
	\fmfstraight
	\fmfleft{l1}
	\fmfright{r1,r2}
	\fmf{photon,tens=2, fore=Dsix}{l1,v1}
	\fmf{plain}{v1,r1}
	\fmf{plain}{v1,r2}
	\fmfv{d.shape=circle, d.size=1.7mm, fore=Dsix,
		lab=\Dsix{$FDF$\hspace{-3.3mm}}, lab.angl=-115, lab.dist=3mm
		}{v1}
	\end{fmfgraph*}};
\draw[->, line width=.7mm] ($(l3.east)+(5mm,0mm)$) -- node[below=0mm, font=\scriptsize]{CCK} ($(r1.west)+(-5mm,0mm)$);
\node [above=9mm of l3.center, anchor=south]{GBAS};
\node [above=9mm of r1.center]{YM};
\end{tikzpicture}
\qquad
\qquad
\begin{tikzpicture}[baseline, yshift=-2mm]
\node (l1) {\begin{fmfgraph*}(7,0)
	\fmfleft{l1}
	\fmfright{r1}
	\fmf{photon, fore=black}{l1,r1}
	\fmfv{d.shape=circle, d.size=1.7mm, fore=black}{r1}
	\fmflabel{\ }{l1}
	\end{fmfgraph*}
	};
\node (r1) [right=1mm of l1]{dim-4 gluon and vertex};
\node (l2) [below=3mm of l1]{\begin{fmfgraph*}(7,0)
	\fmfleft{l1}
	\fmfright{r1}
	\fmf{photon, fore=Dsix}{l1,r1}
	\fmfv{d.shape=circle, d.size=1.7mm, fore=Dsix}{r1}
	\fmflabel{\ }{l1}
	\end{fmfgraph*}
	};
\node (r2) [right=1mm of l2]{dim-6 gluon and vertex};
\node (l5) [above=3mm of l1]{\begin{fmfgraph*}(7,0)
	\fmfleft{l1}
	\fmfright{r1}
	\fmf{plain, fore=black}{l1,r1}
	\fmflabel{\ }{l1}
	\end{fmfgraph*}
	};
\node (r5) [right=1mm of l5]{dim-4 gluon field strength};
\node (l4) [above=3mm of l5]{\begin{fmfgraph*}(7,0)
	\fmfleft{l1}
	\fmfright{r1}
	\fmf{dashes, fore=black}{r1,l1}
	\fmflabel{\ }{l1}
	\end{fmfgraph*}
	};
\node (r4) [right=1mm of l4]{dim-4 bi-adjoint scalar};
\node [right =2mm of r5, yshift=-2mm]{.};
\end{tikzpicture}
\end{equation}

\begin{figure}\centering
\includegraphics[width=0.9\textwidth,trim={0cm 0.4cm 0cm 0cm},clip]{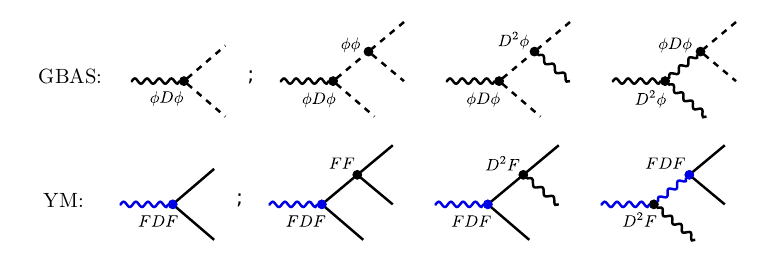}
\caption{Diagrammatic representation of the perturbative solution to the gluon EOM in the YM$+F^3$ theory at three and four points, using the field-strength tensor as an independent function.
Blue lines correspond to EOM solutions at dimension six and end at blue vertices, indicating dimension-six interactions.
}
\label{fig:eom-diag-F3} 
\end{figure}

We emphasize the remarkable fact that the higher-derivative amplitudes of YM$+F^3$ are captured by the GBAS amplitudes \emph{without} higher-derivative interactions.
For example, at three and four points, \autoref{YMF3fromGBAS} is written as
\begin{align}
	\cA_{\textsc{ym},3}^{(6)}
		=&  \frac{1}{\Lambda^2} 
		A_{\phi \phi g}^\textsc{gbas}[12]\, \trF{12}\,,
  \label{YMF3-3pt}
\\\intertext{and}
	\cA_{\textsc{ym},4}^{(6)} = &
	\frac{1%c_6
 }{\Lambda^2}\Big( 
		A_{\phi\phi\phi g}^\textsc{gbas}[123]\,\trF{123}+
		A_{\phi\phi\phi g}^\textsc{gbas}[132]\, \trF{132}
  \nn\\&\;\; +
		A_{\phi\phi g g}^\textsc{gbas}[12]\, \trF{12}+
		A_{\phi g\phi g}^\textsc{gbas}[13]\, \trF{13}+
		A_{g\phi\phi g}^\textsc{gbas}[23]\, \trF{23}
	\Big) \,.
\end{align}
The diagrams that enter the GBAS calculation at these orders and their YM analogues are illustrated in \autoref{fig:eom-diag-F3}.

\subsection{Derivation of YM\texorpdfstring{$+F^3$}{+FFF} numerators}
\label{s:numDim6}

\hyperref[YMF3fromGBAS]{Equation~(\ref*{YMF3fromGBAS})} derives YM$+F^3$ amplitudes from a sum of GBAS amplitudes.
Exactly as in \autoref{s:numDim4}, the transmutation operation can be used to reduce the latter to BAS amplitudes (i.e.\ pure-scalar single-trace tree-level GBAS amplitudes) and hence isolate the BCJ numerators in the trace basis.

For example, at three points, we use \autoref{YMF3-3pt} and a symmetrized version of \autoref{GBAStoBAS-3pt}, namely
\beq 
A^\textsc{gbas}_{\phi \phi g}[12]
= 
\frac{1}{2} A^\textsc{gbas}_{\phi^3}[231]\;
\G[2,3,1]
+\frac{1}{2} A^\textsc{gbas}_{\phi^3}[321]\;
\G[3,2,1]
\,,
\eeq 
to conclude that
\beq 
K^{(6)}[123] = \frac{1}{2\Lambda^2}
\trF{21}G[2,3,1]\,.
\eeq 
The derivation of BCJ numerators at any multiplicity also follows that of~\cite{Cheung:2021zvb}, with an extra symmetrization that relates to the fact that the root leg is a gluon rather than a scalar in the CCK duality at dimension six.
The resulting closed-form expression is
\begin{multline}
	K^{(6)}[12...n] = 
        \frac{1}{\Lambda^2}\sum_{\ell=1}^{n-2}
		\sum_{\tau} \, 
		\frac{1}{|\tau_1|+1} \, 
	\trF{\tau_1\ell}\,
	% \\\times
		\prod_{i=2}^{|\tau|}
		\G\big[(\tau_1...\tau_{i-1})_{<\tau_i},\tau_i,(\tau_1...\tau_{i-1})_{>\tau_i}\ell\big]\,,
\label{numeratorKF3}
\end{multline}
and permutations thereof, with the second sum running over $\tau \in \text{part}(\ell+1,...,n,1,...,\ell-1)$.
This expression relies on the notation of~\cite{Cheung:2021zvb} with small modifications that we discuss now.
The function part$(\sigma)$ is defined as the set of all ordered partitions of the set $\sigma$ into subsets whose elements follow the ordering of $\sigma$.
For example, $1$ should appear on the right of $n$ if both appear in the same subset of a partition.
We also require that the first subset of every partition (i.e.\ $\tau_1$) contains the first element of $\sigma$ but \emph{never} $n$. 
Finally, the greater-than symbol $>$ and less-than symbol $<$ also refer to the ordering
$(\ell+1,...,n,1,...,\ell-1)$.
Namely,
$(\tau_1...\tau_{i-1})_{<\tau_i}$ 
are the elements in 
$\tau_1 \cup ...\cup \tau_{i-1}$ on the left of the 
first element of $\tau_i$ in $(\ell+1,...,n,1,...,\ell-1)$, and $(\tau_1...\tau_{i-1})_{>\tau_i}$ 
are the elements in 
$\tau_1 \cup ...\cup \tau_{i-1}$ on the right of the 
first element of $\tau_i$.

At lowest orders, the part function is
\begin{align}
    \text{part}(23) &= \{2,3\} \nn\\
    \text{part}(234) &= \{ \{23,4\},
                    \{2,34\},
                    \{2,3,4\},
                    \{2,4,3\} \} \nn\\
    \text{part}(341) &= \{ \{31,4\},
                    \{3,41\},
                    \{3,4,1\},
                    \{3,1,4\} \} \,,
\end{align}     
such that \autoref{numeratorKF3} for $n=4$ yields
\begin{align}
	\Lambda^2\, K^{(6)}[1234] &=
		\begin{aligned}[t]
		&\frac{1}{3} \trF{231}\G[23,4,1] 
		+\frac{1}{3} \trF{312}\G[3,4,12]
		\\[1mm]
		+ &\frac{1}{2}\trF{21} \left( \G[2,34,1]
		+ \G[2,3,1] \G[23,4,1]
		+ \G[2,4,1] \G[2,3,41]\right)  
		\\[1mm]
		+&\frac{1}{2} \trF{32} \left(
		 \G[3,41,2]
		+\G[3,4,2] \G[34,1,2]
		+\G[3,1,2] \G[3,4,12]\right)
		\,,
		\end{aligned}
\end{align}
where we remind the reader that 
$\trF{\sigma} \equiv \text{Tr}(\F_{\sigma_1}...\F_{\sigma_{|\sigma|}})$ and $\G$ is defined in \autoref{Gdef}.
We have cross-checked \autoref{YMF3fromGBAS} and \autoref{numeratorKF3} against explicit Feynman diagram calculations in amplitudes with up to seven external particles.

\section{Effective-field-theory extension to dimension eight}
\label{sec:dim8}

The YM$+F^3$ theory of \autoref{eq:F3Lag} does not satisfy the traditional CK duality up to dimension eight, i.e.\ $\cO(1/\Lambda^4)$.
However, the duality can be restored at that order by including a specific dimension-eight interaction \cite{Broedel:2012rc}, resulting in
\begin{align} 
    \mathcal{L}_\textsc{ym}^{(8)} = 
    -\frac{1}{4} F^a_{\mu\nu}F^{a\mu\nu}
	 &- \frac{g}{3 \Lambda^2} 
	f^{abc}\,F^{a\ \nu}_{\ \mu} F^{b\ \rho}_{\ \nu} F^{c\ \mu}_{\ \rho} 
	-\frac{g^2}{4\,\Lambda^4}
	f^{abe}f^{ecd}
	F^a_{\mu\nu}F^b_{\rho\sigma}F^{c\mu\nu}F^{d\rho\sigma}
    + A^{a}_\mu J^{a \, \mu}_{A}
    \,.
    \label{YMdim8Lagr}
\end{align}
In this section, we derive a CCK duality up to $\cO(1/\Lambda^{4})$ between this theory and the following GBAS theory:
\begin{align}
\cL_\textsc{gbas}^{(6)}=&
\cL_{\textsc{ym}}^{(6)}
+\frac{1}{2}D^\mu\phi^{aA}D_\mu\phi^{aA}
-\frac{g\,\lambda}{3}f^{abc}f^{ABC}\phi^{aA}\phi^{bB}\phi^{cC}
+J^{aA}\phi^{aA} 
\nn\\ &\quad 
-\frac{g^2}{4}f^{abe}f^{ecd}\phi^{aA}\phi^{bB}\phi^{cA}\phi^{dB}
-\frac{g}{2\, \Lambda^2}f^{abe}f^{ecd}F_{\mu\nu}^aF^{c\,\mu\nu}\phi^{bA}\phi^{dA}\,,
\label{GBASdim6}
\end{align}
which (except for the $\phi^3$ interaction) results from the dimensional reduction of $\cL_{\textsc{ym}}^{(6)}$ after projection on the massless modes, where the flavors of bi-adjoint scalars correspond to the space-time components of the gauge field along the compact manifold.
This theory therefore satisfies the BCJ relations for all flavor structures~\cite{Chiodaroli:2014xia}, i.e.\ beyond the single-trace order (see also~\cite{Johansson:2013nsa,Chiodaroli:2013upa}).
However, we have dropped all double-trace operators appearing at dimension six in the Lagrangian of \autoref{GBASdim6}, consistently with the EFT power counting of the CCK replacement rule in \autoref{trReplacementF^3}. 
We will show that this rule generalizes to dimension eight, so that the CCK duality combines double-trace dimension-four and single-trace dimension-six GBAS amplitudes to generate pure-gluon dimension-eight amplitudes.
The resulting CCK relation, expressed in terms of one-point functions and using the notation introduced in \autoref{dim6OnePtFctEq}, reads
\beq
\Big[
	   \langle0|A_\mu^a|0\rangle_J^{(6, 1)}
	+ \langle0|A_\mu^a|0\rangle_J^{(4, 2)}
\Big]_\text{GBAS}
\quad
\tikz[baseline]\draw[->, line width=.7mm] (0,.5ex)-- node[below=0mm, font=\scriptsize]{CCK} (1.5cm,.5ex);
\quad
\Big[
	\langle0|A_\mu^a|0\rangle_J^{(8)}
\Big]_\text{YM}
\,.
\label{doubletraceCancellation}
\eeq
The corresponding relation in terms of scattering amplitudes and the explicit treatment of double traces is detailed in \autoref{sec:amplitudesDim8}.

\subsection{Covariant color-kinematics duality between GBAS and YM\texorpdfstring{$+F^3{+}F^4$}{+FFF+FFFF}}

At $\cO(1/\Lambda^2)$, it was found in the previous section that the gluon EOM of the YM EFT can be mapped onto the gluon EOM of the GBAS theory. 
To extend this duality one order higher, we compare the following EOM in the pure-gluon theory at $\cO(1/\Lambda^4)$,
\begin{align}
	&D^\mu F^{a}_{\mu\nu}
	+\frac{g%\, c_6
 }{\Lambda^2} f^{abc} 
	F^{b}_{\mu\rho} D_\nu F^{c\mu\rho} 
	+4 \,\frac{g^2%\, c_6^2
 }{\Lambda^4}
	f^{abe}f^{ecd} F^{c}_{\mu\nu}
	D^\mu F^b_{\rho\sigma} F^{d\rho\sigma}
	= - J^{a}_{\nu} \,,
\label{GluoneomF4}
\end{align}
with the EOMs in the GBAS theory up to $\co(1/\Lambda^2)$,
\begin{align}
	D^\mu  F^{a}_{\mu\nu}
	+& gf^{abc} \phi^{bA} D_\nu \phi^{cA}
		+\frac{g}{\Lambda^2} f^{abc} 
		F^{b}_{\ \mu\rho} D_\nu F^{c\mu\rho}
\nn\\& \hspace{2cm}
			+4\frac{g^2%\,c_6
   }{\Lambda^2} f^{abe}f^{ecd} 
			F^{c}_{\mu\nu} D^\mu \phi^{bA} \phi^{dA}
		= - J^{a}_{\nu} \ ,
\label{GluoneomGBASF3}\\[3mm]
	D^2 \phi^{aA}
	+& \lambda f^{abc}f^{ABC} \phi^{bB} \phi^{cC}
 - {g^2}
		f^{abe}f^{ecd}\phi^{bB}\phi^{cB} \phi^{dA}
  \nn\\& \hspace{3cm}
	- \frac{g^2}{\Lambda^2}
		f^{abe}f^{ecd}F^{b}_{\mu\nu} F^{c\mu\nu} \phi^{dA}
	= J^{aA}\,.
\label{eq:scalareomGBASEFT}
\end{align}
To derive the EOMs in this form, which is suggestive of the CCK duality, we used the lower-order EOMs iteratively in combination with the Jacobi identity.
In particular, the dimension-eight term in \autoref{GluoneomF4} receives contributions from iterations at dimension six, indicating an intricate interplay between different mass dimensions.
We comment further on this point in \autoref{dim8-comment}. 

At the order we are considering, we can decompose the field strength of the pure-gluon theory as $F=F^{(4)}+F^{(6)}+F^{(8)}$, where, as previously, $F^{(d)}$ refers to the field strength solving the gluon EOM at $\cO(1/\Lambda^{d-4})$.
As in \autoref{sec:dim6Details}, we expand $F^{(6)}$ and $F^{(8)}$ in terms of gluons, whereas only $F^{(4)}$ is interpreted as a scalar and taken to evolve through the field-strength EOM of \autoref{fieldstrengtheom}, which we repeat here:
\beq
D^2F^{(4)a}_{\mu\nu}+g\,f^{abc}F^{(4)b}_{\rho[\mu}F^{(4)c\rho}_{\nu]}=-D_{[\mu}J_{\nu]}^a \ .
\eeq
We start by inspecting the dimension-eight term in the pure-gluon EOM in \autoref{GluoneomF4},
\beq
	D^\mu F^{a}_{\mu\nu}
	=-4\, \frac{g^2}{\Lambda^4}
	f^{abe}f^{ecd} F^{c}_{\mu\nu}
	D^\mu F^b_{\rho\sigma} F^{d\rho\sigma}
	+... \,.
\eeq
For solutions up to $\cO(1/\Lambda^4)$, the field strengths on the r.h.s.\ need only satisfy the renormalizable YM EOM.
So, at this order, we can actually solve
\beq
	D^\mu F^{a}_{\mu\nu}
	=-4 \frac{g^2
 }{\Lambda^4}
	f^{abe}f^{ecd}\;
	F^{c}_{\mu\nu}\;
	D^\mu F_{\rho\sigma}^{(4)b}
 \;\:
	F^{(4)d\rho\sigma}
	+... \,,
\label{eq:dim8-pureglue-eom}
\eeq
where $F^{(4)}
$ satisfies the renormalizable YM EOM with source $J$ given in \autoref{fieldstrengtheom}.
We could have added a superscript $(4)$ to the remaining field strength on the r.h.s.\ as well, but as it stands the above has a clear correspondence with the last term ($FD\phi\phi$) of the GBAS EOM in \autoref{GluoneomGBASF3}.
Indeed, when interpreting the flavor structures in terms of Lorentz indices as in $F^a_{\mu\nu}\leftrightarrow \lambda\,\phi^{aA}$,
we know that $F^{(4)}
$ maps to $\phi^{(4)}
$ which solves the GBAS EOM at dimension four with source $DJ$ or, importantly, any EOM like \autoref{eq:scalareomGBASEFT} which reduces to it at $\cO(1/\Lambda^0)$ and in the single-trace sector.
Therefore, the solution to the pure-gluon theory also solves the following EOM,
\beq
	D^\mu F^{a}_{\mu\nu}
	=-4 \frac{g^2 \lambda^2
 }{\Lambda^4}
	f^{abe}f^{ecd}\;
	F^{c}_{\mu\nu}\;
	D^\mu \phi^{bA}
 \;\:
	\phi^{dA}
	+...  
\label{eq:firstTermDim8}
\eeq
This reproduces the last term of the l.h.s.\ of \autoref{GluoneomGBASF3}, up to a factor of $\lambda^2/\Lambda^2$ which we set to one, keeping in mind the CCK rule of \autoref{trReplacementF^3}.
At the diagrammatic level, this implies that any GBAS diagram in which a gluon evolves with this dimension-six $FD\phi\phi$ interaction can be mapped to a dimension-eight diagram in the pure-gluon theory, where the scalar is interpreted as a field strength,
\begin{equation}
\begin{tikzpicture}[baseline]
\node (l2) {\begin{fmfgraph*}(30,20)
	\fmfleft{l1}
	\fmfright{r1,r2,r3}
	\fmf{photon, tens=3, fore=Dsix}{l1,v1}
	\fmf{photon}{v1,r1}
	\fmf{dashes}{v1,r3}
	\fmf{dashes}{v1,r2}
	\fmfv{d.shape=circle, d.size=1.7mm, fore=Dsix,
		lab=\Dsix{$FD\phi\phi$\hspace*{-2mm}}, lab.angl=-116, lab.dist=2.5mm
		}{v1}
	\fmflabel{\ }{l1}
	\end{fmfgraph*}};
\node (r2) [right=2.cm of l2] {\begin{fmfgraph*}(30,20)
	\fmfleft{l1}
	\fmfright{r1,r2,r3}
	\fmf{photon, tens=3, fore=Deight}{l1,v1}
	\fmf{photon}{v1,r1}
	\fmf{plain}{v1,r3}
	\fmf{plain}{v1,r2}
	\fmfv{d.shape=circle, d.size=1.7mm, fore=Deight,
		lab=\Deight{$FDFF$\hspace*{-2.3mm}}, lab.angl=-116, lab.dist=2.8mm
		}{v1}
	\fmflabel{\ }{l1}
	\end{fmfgraph*}};
\draw[->, line width=.7mm] ($(l2.east)+(3mm,0mm)$) -- node[below=0mm, font=\scriptsize]{CCK} ($(r2.west)+(-3mm,0mm)$);
\node [above=2mm of l2]{GBAS};
\node [above=2mm of r2]{YM};
\end{tikzpicture}
\qquad
\begin{tikzpicture}[baseline]
\node (l1) {\begin{fmfgraph*}(7,0)
	\fmfleft{l1}
	\fmfright{r1}
	\fmf{photon, fore=black}{l1,r1}
	\fmfv{d.shape=circle, d.size=1.7mm, fore=black}{r1}
	\fmflabel{\ }{l1}
	\end{fmfgraph*}
	};
\node (r1) [right=1mm of l1]{dim-4 gluon and vertex};
\node (l2) [below=3mm of l1]{\begin{fmfgraph*}(7,0)
	\fmfleft{l1}
	\fmfright{r1}
	\fmf{photon, fore=Dsix}{l1,r1}
	\fmfv{d.shape=circle, d.size=1.7mm, fore=Dsix}{r1}
	\fmflabel{\ }{l1}
	\end{fmfgraph*}
	};
\node (r2) [right=1mm of l2]{dim-6 gluon and vertex};
\node (l3) [below=3mm of l2]{\begin{fmfgraph*}(7,0)
	\fmfleft{l1}
	\fmfright{r1}
	\fmf{photon, fore=Deight}{l1,r1}
	\fmfv{d.shape=circle, d.size=1.7mm, fore=Deight}{r1}
	\fmflabel{\ }{l1}
	\end{fmfgraph*}
	};
\node (r3) [right=1mm of l3]{dim-8 gluon and vertex};
\node (l5) [above=3mm of l1]{\begin{fmfgraph*}(7,0)
	\fmfleft{l1}
	\fmfright{r1}
	\fmf{plain, fore=black}{l1,r1}
	\fmflabel{\ }{l1}
	\end{fmfgraph*}
	};
\node (r5) [right=1mm of l5]{dim-4 gluon field strength};
\node (l4) [above=3mm of l5]{\begin{fmfgraph*}(7,0)
	\fmfleft{l1}
	\fmfright{r1}
	\fmf{dashes, fore=black}{r1,l1}
	\fmflabel{\ }{l1}
	\end{fmfgraph*}
	};
\node (r4) [right=1mm of l4]{dim-4 bi-adjoint scalar};
\node [right =2mm of r1]{.};
\end{tikzpicture}
\label{dim8vertex}
\end{equation}

Besides contributions from this dimension-eight interaction, the solution for $\langle0|A_\mu^a|0\rangle_J$ in the pure-gluon theory also involves diagrams with two dimension-six $F^3$ insertions.
Therefore, the remaining terms in \hyperref[GluoneomF4]{Eqs.\,(\ref*{GluoneomF4}--\ref{eq:scalareomGBASEFT})} need to be compared as well.
However, the $F^3$ interaction of the pure-gluon theory, which leads to a $FDF$ term in the EOM, seems to have two counterparts in the gluon EOM of the GBAS theory, namely
\begin{equation}
    gf^{abc} \phi^{bA} D^\nu \phi^{cA}
    \qquad \text{and} \qquad 
	\frac{g%\, c_6
 }{\Lambda^2} f^{abc} 
		F^{b}_{\ \mu\rho} D^\nu F^{c\mu\rho} \,.
  \label{EOMdegeneracy}
\end{equation}
Two consecutive%
\footnote{Since the diagrams to calculate $\langle0|A_\mu^a|0\rangle_J$ from the EOM are read from left to right, there is a clear ordering in the interactions that occur on the same branch starting from the root leg towards the leaf legs (i.e.\ towards the sources).}
insertions of $FDF$ in the pure-gluon theory have an immediate analog in the GBAS theory.
At the order we consider, the first insertion of $FDF$ can be written as an insertion of $F^{(6)}D^{(4)}F^{(4)}+F^{(4)}D^{(4)}F^{(6)}+F^{(4)}D^{(6)}F^{(4)}$,
where by definition $F^{(6)}$ or $D^{(6)}$ creates the `branch' in the diagram which contains the second $FDF$ interaction.%
\footnote{By $D^{(6)}$, we refer to the piece of the covariant derivative containing a gluon at order $\cO(1/\Lambda^2)$.}
As in \autoref{sec:F3}, this branch is in one-to-one correspondence with a GBAS one where $FDF$ is replaced by $\phi D\phi$.
Therefore, the two $FDF$ insertions in the pure-gluon theory are equivalent in the GBAS theory to an insertion of $FDF$ followed by that of $\phi D\phi$,
\begin{equation}
\begin{tikzpicture}[baseline]
\node (l1) {\begin{fmfgraph*}(30,20)
	\fmfleft{l1}
	\fmfright{r1,r2,r3}
	\fmf{photon, tens=2, fore=Dsix}{l1,v1}
	\fmf{photon}{v1,r1}
	\fmf{phantom}{v1,r3}
	\fmffreeze
	\fmf{photon}{v1,w1}
	\fmf{dashes}{w1,r3}
	\fmffreeze
	\fmf{dashes}{w1,r2}
	\fmfv{d.shape=circle, d.size=1.7mm, fore=black,
		lab=$\phi D\phi$,
		lab.angl=120, lab.dist=1.3mm
		}{w1}
	\fmfv{d.shape=circle, d.size=1.7mm, fore=Dsix,
		lab=$FDF$,
		lab.angl=-110, lab.dist=3.5mm
		}{v1}
	\fmflabel{\ }{l1}
	\end{fmfgraph*}};
\node (r1) [right=2.5cm of l1] {\begin{fmfgraph*}(30,20)
	\fmfleft{l1}
	\fmfright{r1,r2,r3}
	\fmf{photon, tens=2, fore=Deight}{l1,v1}
	\fmf{photon}{v1,r1}
	\fmf{phantom}{v1,r3}
	\fmffreeze
	\fmf{photon, fore=Dsix}{v1,w1}
	\fmf{plain}{w1,r3}
	\fmffreeze
	\fmf{plain}{w1,r2}
	\fmfv{d.shape=circle, d.size=1.7mm, fore=Dsix,
		lab=\Dsix{$FDF$}, lab.angl=120, lab.dist=1.3mm
		}{w1}
	\fmfv{d.shape=circle, d.size=1.7mm, fore=Dsix,
		lab=\Dsix{$FDF$}, lab.angl=-110, lab.dist=3.5mm
		}{v1}
	\fmflabel{\ }{l1}
	\end{fmfgraph*}};
\draw[->, line width=.7mm] ($(l1.east)+(5mm,0mm)$) --node[below=0mm, font=\scriptsize]{CCK} ($(r1.west)+(-5mm,0mm)$);
\node [above=12mm of l1.center]{GBAS};
\node [above=12mm of r1.center]{YM};
\node [right=2mm of r1]{.};
\end{tikzpicture}
\label{eq:two-FDF-onebranch}
\end{equation}

At four points, \hyperref[dim8vertex]{Eqs.~(\ref*{dim8vertex})} and \eqref{eq:two-FDF-onebranch} capture all possibilities and therefore establish a map between the amplitudes.
However, in general, the GBAS theory contains other diagrams involving the following terms of the scalar EOM of \autoref{eq:scalareomGBASEFT},
\begin{equation}
	- \frac{g^2%\,c_6
 }{\Lambda^2}
		f^{abe}f^{ecd}F^{b}_{\mu\nu} F^{c\mu\nu} \phi^{dA}\quad \text{and}\quad
 - {g^2} f^{abe}f^{ecd}\phi^{bB}\phi^{cB} \phi^{dA} \,.
 \label{unwantedOpsScalarEOM}
\end{equation}
These have no direct interpretation in the pure-gluon EOM.
However, we find that the tree amplitudes they give rise to, respectively at the dimension-six and double-trace levels, are related and can cancel each other.
This possibility is suggested by the form of the terms in \autoref{unwantedOpsScalarEOM}.
In the first dimension-six $FF\phi$ term, the field strengths can again be taken to be dimension-four ones $F^{(4)}$ which are equivalent to $\lambda\,\phi$ scalars under the CCK duality.
Up to a factor of $\lambda^2/\Lambda^2$, which we set to one, the two terms therefore become identical.
By including an additional relative sign between the single- and double-trace CCK replacement rules (made explicit in the next section), these two contributions can therefore be canceled against each other,
\begin{equation} \label{GBASscalarLeg1}
\begin{tikzpicture}[baseline]
\node (l1) {\begin{fmfgraph*}(30,20)
	\fmfleft{l1}
	\fmfright{r1,r2,r3}
	\fmf{dashes, fore=Dsix, tens=3}{l1,v1}
	\fmf{dashes}{v1,r1}
	\fmf{plain}{v1,r2}
	\fmf{plain}{v1,r3}
	\fmfv{d.shape=circle, d.size=1.7mm, fore=Dsix,
		lab=\Dsix{$FF\phi$}, lab.angl=113, lab.dist=2.5mm
		}{v1}
	\fmflabel{\ }{l1}
	\end{fmfgraph*}};
\node (l2) [right=2mm of l1] {$+$};
\node (l3) [right=2mm of l2] {\begin{fmfgraph*}(30,20)
	\fmfleft{l1}
	\fmfright{r1,r2,r3}
	\fmf{dashes, fore=black, tens=3}{l1,v1}
	\fmf{dashes}{v1,r1}
	\fmf{dashes}{v1,r2}
	\fmf{dashes}{v1,r3}
	\fmfv{d.shape=circle, d.size=1.7mm, fore=black,
		lab=$\phi\phi\phi$\hspace{-0.5mm}, lab.angl=113, lab.dist=2.5mm
		}{v1}
	\fmflabel{\ }{l1}
	\end{fmfgraph*}};
\hspace{0.5cm}
\node (r1) [right=2.5cm of l3] {\hspace*{0.6cm}$\emptyset$\hspace*{0.6cm}};
\draw[->, line width=.7mm] ($(l3.east)+(5mm,0mm)$) -- node[below=0mm, font=\scriptsize]{CCK} ($(r1.west)+(-5mm,0mm)$);
\node [above=12mm of l2.center]{GBAS};
\node [above=12mm of r1.center]{YM};
\node [right =2mm of r1]{.};
\end{tikzpicture}
\end{equation}

This pattern of cancellations between certain dimension-six single-trace and dimension-four double-trace contributions turns out to be general.
They then also occur in diagrams where a gluon is emitted from a scalar and branches through the term $\frac{g}{\Lambda^2} f^{abc} F^{b}_{\ \mu\rho} D^\nu F^{c,\mu\rho}$ in its EOM.
As seen in \autoref{sec:F3}, this is equivalent to using the term $gf^{abc} \phi^{bA} D^\nu \phi^{cA}$, leading to a double-trace diagram,
\begin{equation}\label{GBASscalarLeg2}
\begin{tikzpicture}[baseline]
\node (l1) {\begin{fmfgraph*}(30,20)
	\fmfleft{l1}
	\fmfright{r1,r2,r3}
	\fmf{dashes, fore=Dsix, tens=2}{l1,v1}
	\fmf{dashes}{v1,r1}
	\fmf{phantom}{v1,r3}
	\fmffreeze
	\fmf{photon,fore=Dsix}{v1,w1}
	\fmf{plain}{w1,r3}
	\fmffreeze
	\fmf{plain}{w1,r2}
	\fmfv{d.shape=circle, d.size=1.7mm, fore=black,
		lab=$D^2\phi$, lab.angl=-120, lab.dist=2mm
		}{v1}
	\fmfv{d.shape=circle, d.size=1.7mm, fore=Dsix,
		lab=\Dsix{$FDF$}, lab.angl=120, lab.dist=1mm
		}{w1}
	\fmflabel{\ }{l1}
	\end{fmfgraph*}};
\node (l2) [right=2mm of l1] {$+$};
\node (l3) [right=2mm of l2] {\begin{fmfgraph*}(30,20)
	\fmfleft{l1}
	\fmfright{r1,r2,r3}
	\fmf{dashes, fore=black, tens=2}{l1,v1}
	\fmf{dashes}{v1,r1}
	\fmf{phantom}{v1,r3}
	\fmffreeze
	\fmf{photon,fore=black}{v1,w1}
	\fmf{dashes}{w1,r3}
	\fmffreeze
	\fmf{dashes}{w1,r2}
	\fmfv{d.shape=circle, d.size=1.7mm, fore=black,
		lab=$D^2\phi$, lab.angl=-120, lab.dist=2mm
		}{v1}
	\fmfv{d.shape=circle, d.size=1.7mm, fore=black,
		lab=$\phi D\phi$, lab.angl=120, lab.dist=1mm
		}{w1}
	\fmflabel{\ }{l1}
	\end{fmfgraph*}};
\hspace{0.5cm}
\node (r1) [right=2.5cm of l3] {\hspace*{0.7cm}$\emptyset$\hspace*{0.7cm}};
\draw[->, line width=.7mm] ($(l3.east)+(5mm,0mm)$) -- node[below=0mm, font=\scriptsize]{CCK} ($(r1.west)+(-5mm,0mm)$);
\node [above=2mm of l2, yshift=8mm]{GBAS};
\node [above=2mm of r1, yshift=8mm]{YM};
\node [right =2mm of r1]{.};
\end{tikzpicture}
\end{equation}

Furthermore, when the interactions appear on different `branches' emerging from the root-leg gluon, all outgoing particles satisfy the dimension-four single-trace EOM at the order that we consider.
The double-trace diagrams then cancel an overcounting that arises from exchanging the distinguishable vertices of the $\phi D\phi$ and $FDF$  interactions of the pure-gluon EOM, leading to an exact equivalence with the pure-gluon diagrams involving a double insertion of the $\frac{g}{\Lambda^2} f^{abc} F^{b}_{\ \mu\rho} D^\nu F^{c,\mu\rho}$ term,
\begin{equation}
\begin{tikzpicture}[baseline]
\node (l1) {\begin{fmfgraph*}(30,20)
	\fmfleft{l0,l1,l2,l3}
	\fmfright{r1,r2,r3,r4}
	\fmf{photon, fore=black, tens=2}{l1,v1}
	\fmf{dashes}{v1,r1}
	\fmf{dashes}{v1,r2}
	\fmf{photon, fore=Dsix, tens=2}{l2,v2}
	\fmf{plain}{v2,r3}
	\fmf{plain}{v2,r4}
	\fmfv{d.shape=circle, d.size=1.7mm, fore=black,
		lab=$\phi D\phi$, lab.angl=-90, lab.dist=3mm
		}{v1}
	\fmfv{d.shape=circle, d.size=1.7mm, fore=Dsix,
		lab=\Dsix{$FDF$}, lab.angl=90, lab.dist=3mm
		}{v2}
	\fmflabel{\ }{l1}
	\end{fmfgraph*}};
\node (l0) [left=0mm of l1] {$2\times$};
\node (l2) [right=2mm of l1] {$+$};
\node (l3) [right=2mm of l2] {\begin{fmfgraph*}(30,20)
	\fmfleft{l0,l1,l2,l3}
	\fmfright{r1,r2,r3,r4}
	\fmf{photon, fore=black, tens=2}{l1,v1}
	\fmf{dashes}{v1,r1}
	\fmf{dashes}{v1,r2}
	\fmf{photon, fore=black, tens=2}{l2,v2}
	\fmf{dashes}{v2,r3}
	\fmf{dashes}{v2,r4}
	\fmfv{d.shape=circle, d.size=1.7mm, fore=black,
		lab=$\phi D\phi$, lab.angl=-90, lab.dist=3mm
		}{v1}
	\fmfv{d.shape=circle, d.size=1.7mm, fore=black,
		lab=$\phi D\phi$, lab.angl=90, lab.dist=3mm
		}{v2}
	\fmflabel{\ }{l1}
	\end{fmfgraph*}};
\node (r1) [right=2.5cm of l3] {\begin{fmfgraph*}(30,20)
	\fmfleft{l0,l1,l2,l3}
	\fmfright{r1,r2,r3,r4}
	\fmf{photon, fore=Dsix, tens=2}{l1,v1}
	\fmf{plain}{v1,r1}
	\fmf{plain}{v1,r2}
	\fmf{photon, fore=Dsix, tens=2}{l2,v2}
	\fmf{plain}{v2,r3}
	\fmf{plain}{v2,r4}
	\fmfv{d.shape=circle, d.size=1.7mm, fore=Dsix,
		lab=\Dsix{$FDF$}, lab.angl=-90, lab.dist=3mm
		}{v1}
	\fmfv{d.shape=circle, d.size=1.7mm, fore=Dsix,
		lab=\Dsix{$FDF$}, lab.angl=90, lab.dist=3mm
		}{v2}
	\fmflabel{\ }{l1}
	\end{fmfgraph*}};
\draw[->, line width=.7mm] ($(l3.east)+(5mm,0mm)$) --node[below=0mm, font=\scriptsize]{CCK} ($(r1.west)+(-5mm,0mm)$);
\node [above=13mm of l2.center]{GBAS};
\node [above=13mm of r1.center]{YM};
\node [right =2mm of r1]{.};
\end{tikzpicture}
\label{eq:two-FDF-twobranches}
\end{equation}
Eventually, using \autoref{doubletraceCancellation} and the appropriate extension of the CCK duality to double traces, one can effectively retain diagrams in which one branch contains first the interaction $\frac{g}{\Lambda^2} f^{abc} F^{b}_{\mu\rho} D^\nu F^{c\mu\rho}$ and then $gf^{abc} \phi^{bA} D^\nu \phi^{cA}$ as in \autoref{eq:two-FDF-onebranch}, as well as diagrams in which the two interactions occur on different branches as in \autoref{eq:two-FDF-twobranches}, without degeneracy.

The different cases discussed above correspond to all possibilities at any multiplicity, proving the validity of our CCK procedure at dimension eight.
For illustration, we display all the five-point diagrams of both the YM+$F^3$+$F^4$ and GBAS theories in \autoref{app:dim8Graphs}.

\subsection{Explicit CCK replacement rules for scattering amplitudes}\label{sec:amplitudesDim8}

As argued above, the CCK duality at dimension eight requires the cancellation of contributions from the $\cO(1/\Lambda^0)$ double-trace sector against some of the $\cO(1/\Lambda^2)$ single-trace ones.
At the level of amplitudes, a relative factor of $-1/\Lambda^2$ is therefore necessary between the single- and double-trace replacements rules,
\begin{equation}
\begin{aligned}
\lambda^{n-2} \, \text{Tr}(T^{A_1}...T^{A_n}) &
\quad
\tikz[baseline]\draw[->, line width=.7mm] (0,.5ex)-- node[below=0mm, font=\scriptsize]{CCK} (1.5cm,.5ex);
\quad
\frac{1}{\Lambda^2} \, \text{Tr}(\F_1...\F_n)
\, ,\\ 
\lambda^{n+m-4} \, \text{Tr}(T^{A_{i_1}}...T^{A_{i_n}}) 
\, \text{Tr}(T^{A_{j_1}}...T^{A_{j_m}})
&
\quad
\tikz[baseline]\draw[->, line width=.7mm] (0,.5ex)-- node[below=0mm, font=\scriptsize]{CCK} (1.5cm,.5ex);
\quad
-\frac{1}{\Lambda^4} \, 
\text{Tr}(\F_{i_1} ...\F_{i_n}) \, 
\text{Tr}(\F_{j_1} ...\F_{j_m})
\,,
\end{aligned}
\label{trReplacementDim8}
\end{equation} 
This generalizes the dimension-six rule of \autoref{trReplacementF^3} to dimension eight and leads to the following formula for YM+$F^3$+$F^4$ amplitudes:
\beq 
\label{YMF4fromGBAS}
\mathcal{A}_{\textsc{ym},n}^{(8)} = 
 \frac{1}{\Lambda^2}
	\sum_{\Phi\in \mathbb{P}^{++}(1...n-1)}
	\;
	\sum_{\sigma\in S(\Phi)/Z_{|\Phi|}}
	A^{(6)}_{\textsc{gbas}}[\sigma]
	\;\trF{\sigma}
 -\frac{1}{\Lambda^4}
	\sum_{\Phi, \bar \Phi }
	\sum_{\sigma, \bar \sigma }
	A^{(4)}_{\textsc{gbas}}[\sigma|\bar\sigma]
	\;\trF{\sigma}
	\;\trF{\bar\sigma}
\,,
\eeq 
where the sums in the second term run over $(\Phi, \bar \Phi)\in \mathbb{P}^{++}(1...n-1)$ with $\Phi \cap \bar \Phi = \emptyset,\Phi<\bar\Phi$ (in some ordering to avoid double counting) and $\sigma\in S(\Phi)/Z_{|\Phi|}$ and similarly for $\bar \sigma$.
In words, these simply span all different double-trace amplitudes with the $n$th particle being a gluon.
It is then relevant to note that the double-trace amplitudes $A^{(4)}_{\textsc{gbas}}[\sigma|\bar\sigma]$ require a minimum of four scalar particles.
Similarly, the amplitudes $A^{(6)}_{\textsc{gbas}}[\sigma]$ are zero when there is only one external gluon.
We have explicitly confirmed \autoref{YMF4fromGBAS} up to six points against Feynman diagram calculations. 

This formula is best exemplified at lowest multiplicities:
{\allowdisplaybreaks%
\begin{align}
    \mathcal{A}_{\textsc{ym},4}^{(8)} &=
        \frac{1}{\Lambda^2} \left( 
        A_{\phi\phi gg}^{(6)}[12]\;\trF{12}
       + A_{\phi g\phi g}^{(6)}[13]\;\trF{13}
        +A_{ g\phi\phi g}^{(6)}[23]\;\trF{23}
        \right)
    \\
\mathcal{A}_{\textsc{ym},5}^{(8)} &= 
      \frac{1}{\Lambda^2} \Bigg( 
        A_{\phi\phi ggg}^{(6)}[12]\;\trF{12}
       + ...
        + A_{ gg\phi \phi g}^{(6)}[34]\;\trF{34}
        \nn\\&\qquad \qquad
        + A_{\phi\phi\phi gg}^{(6)}[123]\;\trF{123}
        + ... 
        +A_{g\phi \phi \phi g}^{(6)}[243]\;\trF{243} \Bigg)
        \label{dim8-5ptExample}\\\nn&\quad 
        -\frac{1}{\Lambda^4}\Bigg( 
            A_{\phi\phi\phi\phi g}^{(4)}[12|34]\;\trF{12}\;\trF{34}+
            ...
            +A_{\phi\phi\phi\phi g}^{(4)}[14|23]\;\trF{14}\;\trF{23}
        \Bigg)\,,
\end{align}}%
where we have suppressed some permutations of the displayed terms, noting again that the $n$th particle is always a gluon.
We also emphasize that the orderings refer to the flavor structures: no color ordering is taken.

\subsection{Dimension eight from dimension four}
\label{s:dim8fromdim4}

Although high-multiplicity expressions become lengthy, the strategy is simple: compute all GBAS amplitudes with $2,...,n-1$ scalars and replace the flavor traces by traces of the linearized field-strength tensors $\F$. 
In fact, we can further leverage a CCK duality, implied by \autoref{GBASscalarLeg1} and \autoref{GBASscalarLeg2}, between dimension-four double-trace GBAS amplitudes and dimension-six single-trace ones. 
After making this duality more precise, we will show that it allows for the derivation of dimension-eight Yang-Mills amplitudes from dimension-four GBAS amplitudes.

Let us consider a single-trace dimension-six GBAS amplitude.
All relevant terms can be found in \autoref{eq:scalareomGBASEFT}.
In particular, the amplitude is computed from 
\beq
D^2 \phi^{aA}
	+ \lambda f^{abc}f^{ABC} \phi^{bB} \phi^{cC}
	- \frac{g^2}{\Lambda^2}
		f^{abe}f^{ecd}F^{b}_{\mu\nu} F^{c\mu\nu} \phi^{dA}
	= J^{aA}\,,
\label{eq:eom-gbas-dual}
\eeq
where, in GBAS theory, the source $J^{aA}$ is independent of the gluon source $J_\mu^a$.
At the order considered, it suffices that the gluon field strength $F$ solves the dimension-four pure-gluon EOM.
Then, CCK for the dimension-four YM theory implies that the EOM of \autoref{eq:eom-gbas-dual} is equivalent to
\beq
\bead
&D^2 \phi^{aA}
	+ \lambda f^{abc}f^{ABC} \phi^{bB} \phi^{cC}
 - \frac{g^2\tilde\lambda^2}{\Lambda^2}
		f^{abe}f^{ecd}\tilde\phi^{b\tilde B}\tilde\phi^{c\tilde B} \phi^{dA}
	= J^{aA}\,,\\
 &D^2\tilde \phi^{a\tilde A} +\tilde\lambda f^{abc}\tilde f^{\tilde A\tilde B\tilde C} \tilde \phi^{b\tilde B} \tilde\phi^{c\tilde C}=\tilde J^{a\tilde A} \ ,
\eead
\eeq
where $\tilde f$ and $\tilde J$ are given by \autoref{eq:sourceRelationGluons} and \autoref{eq:structureConstantRelation}, respectively.
The amplitude which now arises is ``twice single-trace'', i.e.\ it features one trace of $\phi$ flavor and one trace of $\tilde \phi$ flavor.
Now, since $\tilde\phi$ verifies the same EOM as $\phi$, we notice that the diagrams relevant for a given amplitude would precisely be found in the double-trace sector arising from the following EOM,
\beq
D^2 \phi^{aA}
	+ \lambda f^{abc}f^{ABC} \phi^{bB} \phi^{cC}
 - {g^2}
		f^{abe}f^{ecd}\phi^{bB}\phi^{cB} \phi^{dA}
	= J^{aA}\,,
\eeq
which is nothing but the double-trace part of in \autoref{eq:scalareomGBASEFT}.
The resulting CCK duality is such that the replacement rule of 
\autoref{trReplacementF^3} 
should only be applied on the trace that does not involve the root leg.

At the level of the GBAS amplitudes, this implies
\beq 
\label{GBASF2fromGBAS}
\bead
\mathcal{A}_{\textsc{gbas},n}^{(6)}[\Phi\in \mathbb{P}^{++}(1...n)] = 
	&\frac{1}{\Lambda^2}
	\sum_{\scriptsize\begin{matrix}\bar\Phi\in \mathbb{P}^{++}(1...n)\\ \Phi \cap \bar \Phi = \emptyset
	\end{matrix}}
 \,
	\sum_{\bar\sigma\in S(\bar\Phi)/Z_{|\bar\Phi|}}
	A^{(4)}_{\textsc{gbas},n}[\Phi|\bar\sigma]
	\;\trF{\bar\sigma}
\,,
\eead
\eeq
where the amplitudes can be computed from diagrams with any of the scalars in the set $\Phi$ as the root leg.
For example,
\begin{align}
    A^{(6)}_{\phi\phi gg}[12] &= 
    A^{(4)}_{\phi\phi\phi\phi}[12|34]\;\trF{34}
    \\[2mm]
    A^{(6)}_{\phi\phi\phi gg}[123] &=
    A^{(4)}_{\phi\phi\phi\phi\phi}[123|45]\;\trF{45}
    \\[2mm]
    A^{(6)}_{\phi\phi ggg}[12] &=
    A^{(4)}_{\phi\phi\phi\phi g }[12|34]\;\trF{34}+
    A^{(4)}_{\phi\phi\phi g\phi}[12|35]\;\trF{35}+
    A^{(4)}_{\phi\phi g\phi\phi}[12|45]\;\trF{45}\nn\\&\quad 
    +A^{(4)}_{\phi\phi\phi\phi\phi}[12|345]\;\trF{345}+
    A^{(4)}_{\phi\phi\phi\phi\phi}[12|354]\;\trF{354}
\end{align}
where we again emphasize that only flavor orderings are explicitly shown.

Such relations, together with the results of previous sections, lead to 
two new ways of generating $\mathcal{A}^{(8)}_\textsc{ym}$ from GBAS amplitudes.
In the first relation, we conclude that 
any amplitude of the considered YM EFT up to mass dimension eight can be obtained from \emph{renormalizable} GBAS amplitudes using the CCK duality.
The general formula, which we explicitly confirmed through Feynman diagrammatic computations up to six points, reads
\beq 
\label{dim8fromdim4}
\mathcal{A}_{\textsc{ym}}^{(8)} 
 =\frac{1}{\Lambda^4}
\sum_{\Phi, \bar \Phi}
\sum_{\sigma, \bar \sigma}
A^{(4)}_{\textsc{gbas}}[\sigma|\bar\sigma]
  \;\trF{\sigma}
 \;\trF{\bar\sigma}
\,,
\eeq
where the sums run over 
$\Phi, \bar \Phi\in \mathbb{P}^{++}(1...n)$ with
$\Phi \cap \bar \Phi = \emptyset$, $\Phi<\bar\Phi$, 
$\sigma\in S(\Phi)/Z_{|\Phi|}$ and
$\bar\sigma\in S(\bar\Phi)/Z_{|\bar\Phi|}$.
In words, we sum over all different double-trace amplitudes where, in contrast to before, the $n$th particle can be of any type.

The second relation is practically less useful, but conceptually appealing, because it unifies the CCK amplitude relations across mass dimensions.
It makes use of the fact that \autoref{GBASF2fromGBAS} leaves one flavor trace untouched, so it is still true when this trace is replaced according to the dimension-four replacement rule of \autoref{trReplacement}.
This results in the following more symmetric amplitude relation,
\begin{align} \label{unifiedDim8}
    \sum_{\Phi} 
    \left( \mathcal{A}_\Phi^{(4,1)}
    +  \mathcal{A}_{\Phi}^{(4,2)}
    + \mathcal{A}_\Phi^{(6,1)} \right)
\ \ \,
\begin{tikzpicture}[baseline]
\draw[->, line width=.7mm] (0,0ex)-- node[below=0mm, font=\scriptsize]{CCK} (1.5cm,0ex);
\draw[->, line width=.7mm, dash pattern=on 8pt off 3pt] (0,1ex)-- (1.5cm,1ex);
\end{tikzpicture}
\quad \, 
\mathcal{A}_{\textsc{ym},n}^{(4)}+
\mathcal{A}_{\textsc{ym},n}^{(6)}+
\mathcal{A}_{\textsc{ym},n}^{(8)}
\end{align}
where the sum now runs over all subsets of the external particles, including the $n$th one, i.e.\ $\Phi\in \mathbb{P}^{++}(1...n)$.
The superscripts of the GBAS amplitudes refer to mass dimension and number of flavor trace factors, respectively, while their subscripts indicate which external particles are scalars (all others being gluons).
Following the CCK replacement rule, flavor traces are replaced according to 
$ \lambda^{|\sigma|-2}\,\text{Tr}(\sigma)
    \to {\trF{\sigma}}/{\Lambda^2}$ 
if the root leg is a gluon ($n \notin \sigma$) and according to 
$ \lambda^{|\sigma|-1}\,\text{Tr}(\sigma n)
    \to \trFt{\sigma \tilde n}$ if it is a scalar ($n\in\sigma$),
with an overall minus sign for the double-trace contribution.

It is now tempting to speculate that the relations between the GBAS and YM theories extend to even higher orders in their EFT expansions, although the cancellations between single- and higher-trace are not a priori obvious.
We explore this in the next section.

\subsection{Derivation of YM\texorpdfstring{$+F^3{+}F^4$}{+FFF+FFFF} numerators}
As done in \hyperref[s:numDim4]{Sections~}\ref{s:numDim4} and \ref{s:numDim6}, the r.h.s.\ of \autoref{dim8fromdim4} can also be expressed in terms of (single-trace) BAS amplitudes, allowing for a derivation of the BCJ numerators.
In fact, with the closed form of Yang-Mills numerators at hand \cite{Cheung:2021zvb}, the procedure is straightforward. 
Starting from the dimension-eight amplitude,
\beq 
 \mathcal{A}_{\textsc{ym}}^{(8)} =
     \frac{1}{\Lambda^4} \left( 
A_{\phi\phi \phi\phi}^{(4)}[12|34]
\;\trF{12}\;\trF{34}
+
A_{\phi\phi \phi\phi}^{(4)}[13|24]
\;\trF{13}\;\trF{24}
+
A_{\phi\phi \phi\phi}^{(4)}[14|23]
\;\trF{14}\;\trF{23}
      \right)\,,
\eeq 
and using the fact that \cite{Cheung:2017ems}
\beq 
    A_{\phi\phi \phi\phi}^{(4)}[12|34]
    = 4\,
     \partial_{\epsilon_1\cdot\epsilon_2}
     \partial_{\epsilon_3\cdot\epsilon_4} \, 
     \mathcal{A}_{\textsc{ym},4}^{(4)}\,,
\eeq 
it follows that
\beq 
    K^{(8)}[1234] = 
    \frac{4}{\Lambda^4}
    \begin{aligned}[t]
\Big(\,&   \trF{12}\trF{34}\,
    \partial_{\epsilon_1\cdot\epsilon_2}
     \partial_{\epsilon_3\cdot\epsilon_4}
\\+&
    \trF{13}\trF{24}\,
    \partial_{\epsilon_1\cdot\epsilon_3}
     \partial_{\epsilon_2\cdot\epsilon_4}
\\+&
    \trF{14}\trF{23}\,
    \partial_{\epsilon_1\cdot\epsilon_4}
     \partial_{\epsilon_2\cdot\epsilon_3}
\Big) K^{(4)}[1234]\,.
\end{aligned}
\eeq 
We leave the derivation of a closed form formula for arbitrary multiplicity at dimension eight for future work.

\subsection{Comments on restricting to dimension eight only}\label{dim8-comment}

From \autoref{dim8vertex}, and the associated study at the level of the EOMs, it might seem that the CCK duality can be applied separately to the dimension-eight vertex, even though this vertex does not satisfy the traditional CK duality by itself.
It is however important to realize that the dimension-eight interaction in the EOM of \autoref{GluoneomF4} is not in one-to-one correspondence with the dimension-eight operator in the Lagrangian of \autoref{YMdim8Lagr}.
Instead, iterations of the dimension-six terms in the EOM are necessary to bring the interaction in this form.
It would therefore not be consistent to consider the dimension-eight term separately at the level of the EOM. 
This suggests that the traditional CK duality is necessary for the CCK duality.

\section{Effective-field-theory extension beyond dimension eight}
\label{sec:beyonddim8}

The EFT analysis above suggests that gluon amplitudes at higher mass dimensions can be obtained from lower-order GBAS amplitudes using the CCK duality.
This provides a map from the GBAS EFT into the YM EFT, where both theories consist of a tower of operators that satisfy the traditional CK duality.
It was previously found that such CK-dual towers of operators are encoded by the so-called $(DF)^2$+YM and $(DF)^2$+YM+$\phi^3$ theories~\cite{Johansson:2017srf, Azevedo:2018dgo}.
Their double copies were also been studied in~\cite{Johansson:2018ues,Ben-Shahar:2021zww}.
In this section, we explore the correspondence between these theories and the EFTs considered above, as well as the CCK duality between them.

\paragraph{$(DF)^2$+YM.}
In four space-time dimensions, the $(DF)^2$+YM Lagrangian can be written as~\cite{Johansson:2017srf}
\begin{align}
    \mathcal{L}_{(DF)^2+\text{YM}} = &
        \ -\frac{1}{4}(F_{\mu\nu}^a)^2 
        + \frac{1}{2\,m^2}(D^\mu F^a_{\mu\nu})^2 
        + \frac{1}{2}(D_\mu \varphi^\alpha)^2
        -\frac{m^2}{2}  (\varphi^\alpha)^2 
        \nn\\[1mm] &        
        + \frac{m\,g}{3!} d^{\alpha\beta\gamma} \varphi^\alpha\varphi^\beta\varphi^\gamma
        + \frac{g}{2\,m} C^{\alpha ab} \varphi^\alpha F^a_{\mu\nu}F^{b\mu\nu} 
        - \frac{g}{3\,m^2} f^{abc} F_{\ \mu}^{a\ \nu}F_{\ \nu}^{b\ \rho}F_{\ \rho}^{c\ \mu}
        \,,
\end{align}
where $\varphi^\alpha$ is a real scalar with mass $m$ in a real representation of the $\text{SU}(N)$ gauge group. 
The Clebsch-Gordan coefficients $C^{\alpha ab}$ and $d^{\alpha \beta \gamma}$ satisfy the following relations~\cite{Johansson:2017srf},
\begin{align}
    C^{\alpha ab} C^{\alpha cd} &= f^{ace}f^{edb} + (c\leftrightarrow d) \\
    C^{\alpha ab} d^{\alpha \beta \gamma} &= (T^a)^{\beta \alpha}(T^b)^{\alpha\gamma} + C^{\beta ac}C^{\gamma cb} + (a\leftrightarrow b)
    \,,
\end{align}
where $(T^a)^{\alpha \beta}$ are the generators of the representation of $\varphi^\alpha$.
The $(DF)^2$ term gives corrections to the gluon propagator, which (after gauge fixing) can be written as 
\begin{equation}
\begin{tikzpicture}[baseline]
\node (r1) {\begin{fmfgraph*}(20,15)
	\fmfleft{l1}
	\fmfright{r1,r2}
	\fmf{photon,tens=0.6, lab=$p$}{l1,v1}
	\fmf{phantom}{v1,r1}
	\fmf{phantom}{v1,r2}
    \fmfv{lab=$\mu$}{l1}
	\fmfv{d.shape=circle, d.size=1.7mm, fore=black,lab=$\nu$
		}{v1}
	\end{fmfgraph*}};
\end{tikzpicture}
\hspace{0.5mm}
= \quad \frac{-i \, \eta_{\mu\nu}}{p^2-\frac{p^4}{m^2}}
\quad = \quad 
-i\,\eta_{\mu\nu} \left( \frac{1}{p^2} - \frac{1}{p^2-m^2} \right) 
\,,
\label{eq:p4GluonPropagator}
\end{equation}
indicating the presence of a ghost of mass $m$.
It was found in~\cite{Johansson:2017srf} that the $(DF)^2$+YM theory satisfies the traditional CK duality at tree level for any value of the mass $m$.

To compare with the dimension-eight YM Lagrangian of \autoref{YMdim8Lagr}, we take the heavy-mass limit and integrate out the scalar at tree level by replacing it recursively by its classical solution, which solves the EOM,
\begin{align}
    \varphi^\alpha_\text{cl} = \ &  
     \frac{g}{2\,m^3} \, C^{\alpha ab} F^a_{\mu\nu}F^{b\mu\nu}
    -\frac{g}{2\,m^5}\, C^{\alpha ab} D^2 (F^a_{\mu\nu}F^{b\mu\nu})
    +\cO\!\left(1/m^7 \right)\,.
\end{align}
This yields the EFT Lagrangian
\begin{align}
    \mathcal{L}_{(DF)^2+\text{YM}}^\text{EFT} 
    \stackrel{\textsc{fr}}{=} \ & \ 
            -\frac{1}{4}(F_{\mu\nu}^a)^2 
        -  \frac{g}{3\,m^2} f^{abc} F_{\ \mu}^{a\ \nu}F_{\ \nu}^{b\ \rho}F_{\ \rho}^{c\ \mu}
        - \frac{g^2}{4\,m^4}\, f^{abe}f^{ecd} F^a_{\mu\nu} F^b_{\rho\sigma} F^{c\mu\nu} F^{d\rho\sigma} 
        \nn\\&
        -\frac{g^2}{m^6}\,
        f^{abe}f^{ecd} 
        F^a_{\mu\nu} 
        D_\tau F^b_{\rho\sigma} D^\tau F^{c\mu\nu}  
        F^{d\rho\sigma}
        +\cO\!\left(1/m^8\right)
    \,. 
    \label{DF2YMEFT}
\end{align}
We emphasized that we have also performed a field redefinition (FR) in order to replace $(DF)^2$ by operators involving more fields and higher mass dimensions.
Indeed, $(DF)^2$ can be treated perturbatively in the EFT limit of small $1/m$.
In other words, we integrate out the massive ghost at tree level.
Besides exhibiting the correspondence with \autoref{YMdim8Lagr}, the above Lagrangian includes an operator satisfying the CK duality at the next order in $1/m$.
This is a natural candidate operator for the CCK duality as well.

\paragraph{$(DF)^2$+YM+$\phi^3$.}
The $(DF)^2$+YM+$\phi^3$ theory is defined by the Lagrangian
\begin{align}
    \mathcal{L}_{(DF)^2+\text{YM}+\phi^3} = &
    \mathcal{L}_{(DF)^2+\text{YM}} 
        + \frac{1}{2} (D_\mu \phi^{aA})^2
        - \frac{g\,\lambda}{3} f^{abc}f^{ABC} \phi^{aA}\phi^{bB}\phi^{cC}
        + \frac{m\,g}{2} C^{\alpha ab} \varphi^\alpha \phi^{aA}\phi^{bA} 
        \,,
        \label{YMDF2EFT}
\end{align}
and also satisfies the traditional CK duality at tree level~\cite{Johansson:2017srf}.  
As before, the heavy scalar can be integrated out, to give the EFT Lagrangian
\begin{align}
    \mathcal{L}_{(DF)^2+\text{YM}+\phi^3}^\text{EFT}
&    \stackrel{\textsc{fr}}{=}  
\mathcal{L}_{(DF)^2+\text{YM}}^\text{EFT}
        + \frac{1}{2} (D_\mu \phi^{aA})^2
        \nn\\&- \frac{g \,\lambda}{3}  \,f^{abc}f^{ABC} \phi^{aA}\phi^{bB}\phi^{cC}
-\frac{g^2}{4}f^{abe}f^{ecd}\phi^{aA}\phi^{bB}\phi^{cA}\phi^{dB}
        \nn\\&
-\frac{g}{2\, m^2}f^{abe}f^{ecd}F_{\mu\nu}^aF^{c\,\mu\nu}\phi^{bA}\phi^{dA}
        -\frac{g^2}{m^2}
        f^{abe}f^{ecd} 
        \phi^{aA} D_\mu\phi^{bB} 
         D_\mu\phi^{cA} \phi^{dB}
        \nn\\&
        -2\frac{g^2}{m^4} f^{abe}f^{ecd} 
        F^a_{\mu\nu} D_\rho \phi^{bA}
        D_\rho F^{c\mu\nu} \phi^{dA}
        +\cO\!\(1/m^6\)
        \,,
\end{align}
where we have neglected all terms that contribute beyond the dimension-six double-trace and dimension-eight single-trace orders, because these only contribute to YM amplitudes of dimension twelve and higher after application of the CCK replacement rule and are therefore not relevant to map to \autoref{DF2YMEFT}.
This Lagrangian follows from the dimensional reduction of \autoref{DF2YMEFT}.%
\footnote{As before, the $\phi^3$ vertex does not follow from dimensional reduction but needs to be included by hand.
An ambiguity could arise if it were to matter whether this interaction would be included before or after performing field redefinitions.
However, we find that the difference between these two treatments is a term of the form $ f^{abx}f^{ycd}f^{yex}f^{ABC} \phi^{aA}\phi^{bB}\phi^{cC} F^d_{\mu\nu} F^{e\mu\nu}$, which vanishes due to the Jacobi identity.}

How can these two massive theories be related by the CCK duality? At large $m$, they generate the EFTs we encountered before and extend them to arbitrary mass dimension.
Due to our power counting (see e.g.\ \autoref{trReplacementDim8}), we expect that an increasing number of flavor traces be needed in the CCK replacement rule when considering higher-and-higher EFT orders.
To relate the $(DF)^2+$YM and $(DF)^2+$YM$+\phi^3$ amplitudes for general $m$, the treatment of an arbitrary number of trace factors should therefore be needed.
Deriving the associated complete set of rules is however beyond the scope of this paper.
Nevertheless, restricting to low-multiplicity amplitudes, we can test the CCK duality at the level of these two massive theories without having to treat large numbers of flavor traces.
In particular, up to six points, the amplitudes featuring at least one external gluon do not involve triple flavor traces.
Therefore, the CCK map already derived could potentially extend to all orders in the EFT expansion at that multiplicity.

We have indeed explicitly confirmed that the CCK replacement rule  of \autoref{trReplacementDim8} maps $(DF)^2$+YM+$\phi^3$ amplitudes to $(DF)^2$+YM ones for \emph{any} value of the mass $m$.%
\footnote{As we are interested in comparing the EFTs of these theories, we did not consider amplitudes with external heavy scalars $\varphi$.
We leave the discussion of such amplitudes to future work.} 
The corresponding formula reads
    \beq
    \cA_{(DF^2)+\text{YM}} 
    - \mathcal{A}_{\textsc{ym}}^{(4)}
    =
    \sum_{\Phi,\sigma} 
    A_{(DF^2)+\text{YM}+\phi^3}[\sigma]
    \frac{\trF{\sigma}}{m^2}
    -
    \sum_{\Phi,\sigma,\bar\Phi,\bar\sigma} 
    A_{(DF^2)+\text{YM}+\phi^3}[\sigma|\bar\sigma]
    \frac{\trF{\sigma}}{m^2}
    \frac{\trF{\bar\sigma}}{m^2} \ ,
    \label{generalmRelation}
    \eeq
which is valid for $n\leq 6$ and where the sums are taken with a root-leg gluon as in \autoref{YMF4fromGBAS}.
We remind the reader of the fact that $\lambda$ has been implicitly set to $1$ on the r.h.s..
This result implies that the CCK relations extend to all orders in the EFT expansion, up to six-point amplitudes at least.

Beyond six points, we expect that \autoref{generalmRelation} would receive triple-trace contributions. 
Indeed, at seven points, we have confirmed that the dimension-ten EFT amplitudes generated from $\mathcal{L}_{(DF)^2+\text{YM}+\phi^3}^\text{EFT}$ and $\mathcal{L}_{(DF)^2+\text{YM}}^\text{EFT}$ are related by the following generalization of \autoref{trReplacementDim8}: 
\begin{equation}\label{trReplacementAllDim}
    \prod_{i=1}^{n} \lambda^{|\sigma_i|-2}\,
                      \text{Tr}(\sigma_i)
\quad
\tikz[baseline]\draw[->, line width=.7mm] (0,.5ex)-- node[below=0mm, font=\scriptsize]{CCK} (1.5cm,.5ex);
\quad
    (-1)^{n+1} \prod_{i=1}^{n} \frac{\trF{\sigma_i}}{m^2}\,,
\end{equation}  
for products of $n$ traces.
We conjecture that this CCK replacement rule is valid at all mass dimensions and multiplicities.

It is then also possible to generalize the unified treatment of all amplitudes, regardless of their $n$th root leg, beyond the dimension-eight relation of \autoref{unifiedDim8}.
Schematically, one obtains
\beq \label{generalmRelationUnified}
    \sum_\text{all permutations}
    \mathcal{A}_{(DF^2)+\text{YM}+\phi^3,n}
\ \ \,
\begin{tikzpicture}[baseline]
\draw[->, line width=.7mm] (0,0ex)-- node[below=0mm, font=\scriptsize]{CCK} (1.5cm,0ex);
\draw[->, line width=.7mm, dash pattern=on 8pt off 3pt] (0,1ex)-- (1.5cm,1ex);
\end{tikzpicture}
\quad \, 
\cA_{(DF^2)+\text{YM},n} 
\eeq
where the sum runs over the permutations of all $(DF^2)+\text{YM}+\phi^3$ amplitudes with $2\leq m \leq n$ external scalars $\phi$.
The replacement rule for flavor traces is $ \lambda^{|\sigma|-2}\,\text{Tr}(\sigma) \to {\trF{\sigma}}/{m^2}$ if the root leg is a gluon ($n \notin \sigma$) and $ \lambda^{|\sigma|-1}\,\text{Tr}(\sigma n) \to \trFt{\sigma \tilde n}$ if it is a scalar ($n\in\sigma$), with a crucial minus sign for multiple traces as in \autoref{trReplacementAllDim}.
This relies on the CCK duality between GBAS amplitudes of different mass dimensions, which we exploited already in \autoref{s:dim8fromdim4} at dimension eight.
We have explicitly confirmed \autoref{generalmRelationUnified} up to six points.

Finally, we have tested the extension of 
\autoref{dim8fromdim4}, which relates dimension-four GBAS to dimension-eight YM amplitudes, to the full tower of EFT operators.
We find that such a relation does indeed hold for general $m$ up to at least six points.
At six points, beyond dimension eight, there are contributions from triple-trace amplitudes, because \autoref{dim8fromdim4} does not require a root-leg gluon. 
These are captured by the formula,
\begin{align}
    \cA_{(DF^2)+\text{YM}} 
    - \mathcal{A}_{\textsc{ym}}^{(4)}
    - \mathcal{A}_{\textsc{ym}}^{(6)} 
    =&
    \frac{1}{m^4}\sum_{\Phi,\bar \Phi} \sum_{\sigma, \bar \sigma} 
    A_{(DF^2)+\text{YM}+\phi^3}[\sigma|\bar \sigma] 
    \;\trF{\sigma}
    \;\trF{\bar \sigma}
    \nn\\&
    -\frac{2}{m^6}
    \sum_{\Phi_1,\Phi_2,\Phi_3} 
    A_{(DF^2)+\text{YM}+\phi^3}[\Phi_1|\Phi_2|\Phi_3]
    \;\trF{\Phi_1}
    \;\trF{\Phi_2}
    \;\trF{\Phi_3}
    \label{generalmRelation2}
\end{align}
where the sums in the first line are the same as in \autoref{dim8fromdim4}, while the triple trace sums satisfy 
$\Phi_1,\Phi_2,\Phi_3 \in \mathbb{P}^{++}(1...6)$ with 
$\Phi_i \cap \Phi_j = \emptyset$ and  
$\Phi_1<\Phi_2<\Phi_3$, referring again to some ordering to avoid overcounting.
The fact that the $(DF)^2$+YM amplitudes can be decomposed in multiple different ways, namely according to 
\hyperref[generalmRelation]{Eqs.~(\ref*{generalmRelation})}, 
\eqref{generalmRelationUnified}
and \eqref{generalmRelation2}, requires an intricate self-duality of the $(DF)^2$+YM+$\phi^3$ amplitudes which deserves to be better understood.
Comparing the EOMs of these theories for general mass would certainly shed light on this mapping and clarify how to extend it.
We leave such explorations to future work.

\section{Conclusions}
\label{sec:conclusions}

We extend the CCK duality between the GBAS and YM theories to the higher-derivative regime, focusing on EFTs which satisfy the traditional color-kinematics duality.
We first investigate the operators of lowest mass dimensions, before proposing generalizations to an infinite tower of operators.
We find that a pure-gluon theory with operators of mass dimension $\leq 2k$ can be generated from a theory of gluons and bi-adjoint scalars featuring operators of mass dimension $\leq 2(k-1)$.
Remarkably, the latter GBAS EFT can be obtained from the dimensional reduction of a pure-gluon theory including operators of mass dimension $\leq 2(k-1)$, up to a cubic scalar interaction added a posteriori.

The established higher-derivative CCK duality follows from a clear correspondence between all terms in the equations of motion of the two theories.
Expressed at the level of amplitudes, it proceeds through the replacement of traces of scalar flavor structures by kinematic factors.
Remarkably, the CCK duality requires intricate cancellations between GBAS amplitudes featuring different numbers of flavor traces.
These for instance occur between the double-trace dimension-four and single-trace dimension-six GBAS sectors needed to generate dimension-eight YM amplitudes.
With this insight, we establish simple all-multiplicity relations between GBAS and YM amplitudes up to dimension eight. 

Specifically, dimension-six YM+$F^3$ amplitudes are derived from single-trace dimension-four GBAS amplitudes through the replacement of flavor traces by local gauge-invariant functions of the kinematics.
We leverage this relation to derive closed-form expressions for the BCJ numerators of the dimension-six YM theory, at any multiplicity and with manifest gauge invariance on all legs.

Beyond the minimal higher-derivative correction, we focus on towers of EFT operators which are compatible with the traditional CK duality.
We find two ways of constructing the CK-dual dimension-eight YM amplitudes from GBAS inputs.
Firstly, the same CCK replacement rule, applied to dimension-four double-trace---with a crucial minus sign---and dimension-six single-trace GBAS amplitudes, results precisely in pure-gluon amplitudes.
Alternatively, the dimension-eight YM amplitudes are derived from dimension-four double-trace GBAS ones only, when more permutations of the external particles are included.
These relations again lead to a simple procedure to construct the BCJ numerators of the YM theory at dimension eight.
This serves as a new proof for the standard CK duality up to dimension eight, at any multiplicity and tree level.
By confirming our CCK relations in the $(DF)^2$+YM\,(+$\phi^3$) theories at low multiplicity, we have obtained strong evidence that they extend to all orders in the EFT expansion.

Several directions would deserve to be investigated further.
Relations conjectured between the $(DF)^2+$YM$(+\phi^3)$ theories at all EFT orders and multiplicities ought to be (dis)proved.
We find evidence that the existence of a CK duality is sufficient to ensure a CCK one, but we do not know whether it is a necessary condition.
Hence, it would be very interesting to explore further the equations of motion generated by gluon operators which are not those considered in this work.
Similarly, staying in the realm of CK-dual theories, there may be BCJ-compatible operators at high mass dimensions beyond those encoded in $(DF)^2+$YM theory. 
If so, understanding how they enter a CCK duality would be insightful.

Beyond the ideas touched upon in this paper, we have not studied the double copy to gravity.
We have derived BCJ numerators for the YM EFT up to dimension eight, which can directly be used in the traditional double copy, but it would be a natural extension of our work to make manifest a CCK duality for higher-derivative corrections to gravity, as done in~\cite{Cheung:2021zvb} at dimension four.
This reference also identified a CCK duality at the level of the EOMs of the BAS and NLSM theories, which is another direction of investigation that we plan to explore in the future.
We anticipate that the NLSM$+\phi$ theories found in~\cite{Cachazo:2016njl,Carrasco:2016ygv} are likely to play an important role.
Another insight of~\cite{Cheung:2021zvb} which we have not extended yet to EFTs concerns the relation between conserved currents, which is possibly affected by our enlarged CCK dualities.
Finally, the ultimate amplitude relations that we find are simpler than could be expected from a first inspection of the equations of motion, due to intricate cancellations between multi-trace replacements.
It would be worth exploring whether such relations extend to the level of loop integrands.
The same applies to the CCK duality more generally.

\section*{Acknowledgments}

The calculations for this work were performed using FORM \cite{Vermaseren:2000nd, Ruijl:2017dtg, Ueda:2020wqk}.
We thank Christophe Grojean and Camila Machado for their collaboration at the early stages of this work, and Clifford Cheung for comments on the paper. 
JRN thanks Felipe D\'iaz-Jaramillo for insightful discussions.
QB is supported by the Office of High Energy Physics of the U.S.\ Department of Energy under Contract No.\ DE-AC02-05CH11231.
GD is a Research Associate of the the Fund for Scientific Research~--~FNRS.
JRN is supported by the Deutsche Forschungsgemeinschaft (DFG, German Research Foundation) -- Projektnummer 417533893/GRK2575 ``Rethinking Quantum Field Theory''.

%%%%%%%%%%%%%%%%%%%%%%%%%%%%%%%%%%%%%%%%%%%%%%%%%%%%%%%%%%%%
\newpage 
\appendix

%%%%%%%%%%%%%%%%%%%%%%%%%%%%%%%%%%%%%%%%%%%%%%%%%%%%%%%%%%%%
\section{Diagrammatic map at five points and dimension eight}
\label{app:dim8Graphs}

In this appendix we exemplify the CCK duality between the GBAS theory at mass dimension four (double trace) and six (single trace) and the YM theory at mass dimension eight.
Blue and green lines correspond to EOM solutions at dimension six and eight, respectively. 

\vspace{0.6cm}

\begin{figure}[h!]\centering
	\adjustbox{max width=\textwidth}{%
	\includegraphics[trim={2.5cm 3cm 2.8cm 5cm}, clip]{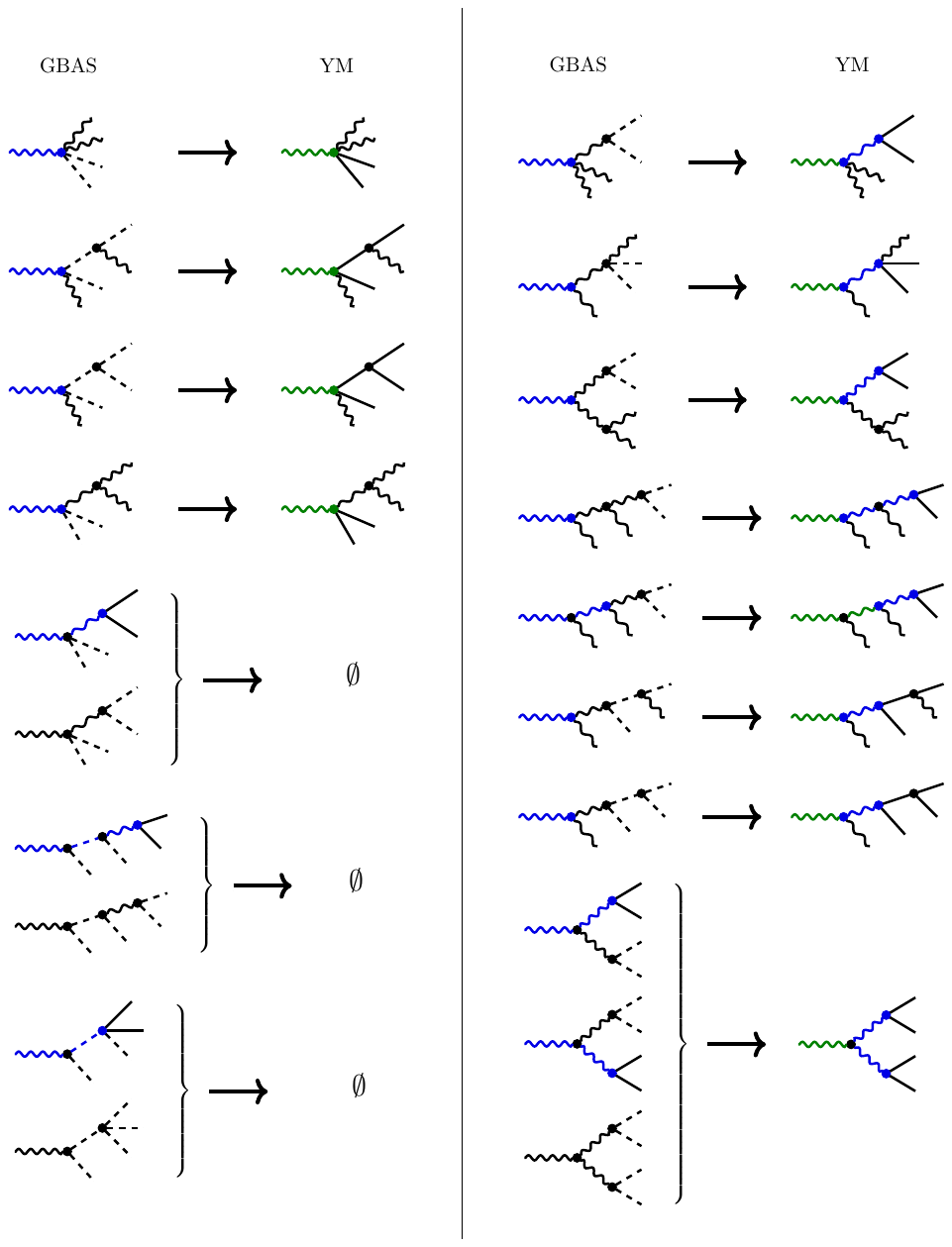}%
	}
\end{figure}
\newpage

\end{fmffile}

\bibliographystyle{apsrev4-1_title}
\bibliography{biblio}
\end{document}